\documentclass[12pt,authoryear]{elsarticle}
 
\usepackage{subfigure}
\usepackage[colorlinks]{hyperref} 

\usepackage{amssymb}
\usepackage{amsfonts}
\usepackage{dsfont}
\usepackage{amsmath}
\usepackage{multirow}

\usepackage{bm}
\usepackage{xcolor}
\usepackage{siunitx}

\newcommand{\be}{\begin{equation}}
\newcommand{\ee}{\end{equation}}
\newcommand{\bq}{\begin{eqnarray}}
\newcommand{\eq}{\end{eqnarray}}

\newcommand{\bc}{\begin{center}}
\newcommand{\ec}{\end{center}}
\newcommand{\beq}{\begin{equation}}
\newcommand{\eeq}{\end{equation}}
\newcommand{\bea}{\begin{eqnarray}}
\newcommand{\eea}{\end{eqnarray}}

\newcommand  {\TD} {\mathrm{TD}}
\newcommand  {\sub} {\mathrm{sub}}
\newcommand  {\supp} {\mathrm{sup}}
\newcommand  {\laser} {\mathrm{laser}}
\newcommand {\at} {\mathrm{at}}
\newcommand {\scat} {\mathrm{sc}}
\newcommand {\sat} {\mathrm{sat}}
\newcommand {\pair} {\mathrm{pair}}

\newcommand  {\twothree}    {$|2\rangle \rightarrow |3'\rangle$ }
\newcommand {\inphyni} {Universit\'e C\^ote d'Azur, CNRS, Institut de Physique de Nice, France}

\journal{Advances in Atomic, Molecular and Optical Physics}

\begin{document}

\begin{frontmatter}

\title{Super- and subradiance in dilute disordered cold atomic samples: observations and interpretations}

\author{William Guerin}
\address{\inphyni}

\begin{abstract}
When a photon is sent onto an atomic ensemble, it interacts collectively with the $N$ atoms of the sample and not simply with one of them. This results in measurable modifications in the scattering rate, the emission diagram or the temporal dynamics. Among these collective effects, here, we study in detail the temporal dynamics of the decay of the scattered light after switching off the driving laser. Right after the switch-off, the decay can be faster than for a single atom, whereas at later time, the decay becomes slower. We refer to this behavior as superradiance and subradiance, respectively. We present in this Chapter our investigations of super- and subradiance in disordered cold atomic ensembles, mainly with low-density samples (dilute regime) and a weak excitation (linear-optics regime), but we also studied the first corrections beyond these limiting cases. We emphasize the different possible interpretations of these phenomena, with in particular the presentation of an optical description of these effects, which sheds a new light on cooperative scattering in disordered samples and provides a more intuitive understanding of the physical mechanisms at play. 
\end{abstract}


\end{frontmatter}

\today

\maketitle

\tableofcontents




\ \\

In this Chapter, we present and summarize our recent work on superradiance and subradiance in disordered cold atomic samples. Most of this work deals with the case of a dilute sample, when the average distance between atoms is much larger than the wavelength. It also mostly concerns the weak-driving limit, i.e. the linear-optics regime. The results presented in this Chapter are reported in more details in several peer-reviewed articles published between 2016 and 2022. However, our understanding of these phenomena has evolved over time and we feel that a self-consistent presentation of the whole work may be useful to the community. This is the motivation behind the writing of this Chapter. It also contains one new unpublished result, Fig.\,\ref{fig.CDvsRW}.

We start with a short introduction to set the context of our work. Then we present our numerical and experimental observations of subradiance (section \ref{sec.subradiance}), and in section \ref{sec.superradiance} our observations of superradiance. In section \ref{sec.atomic} we discuss the interpretation of super- and subradiance in terms of collective atomic modes, which is the most common interpretation. In section \ref{sec.optical} we present an alternative picture, based on optical phenomena such as dispersion and multiple scattering. We argue that this optical interpretation is useful to physically understand a lot of properties of super- and subradiance. Finally, in Section \ref{sec.beyond} we present a few results that go beyond the limits of dilute sample and linear-optics regime, before concluding.

\section{Short introduction on super- and subradiance}\label{sec.intro}

The topic of superradiance is so large that any attempt to do an exhaustive review would fail. We thus give a brief and very incomplete overview of the past work on the subject\footnote{In particular we say nothing about the topics of superradiant lasing \citep{Bohnet:2012}, Dicke model \citep{Roses:2020, Ferioli:2022}, and black-hole superradiance \citep{Brito:2020}.}, with slightly more details on the more recent, so-called `single-photon superradiance'.

\subsection{Dicke's superradiance}

In 1954 R.\,H.\,Dicke published an article entitled \textit{``Coherence in spontaneous radiation processes''} \citep{Dicke:1954}, which introduced the concept of superradiance. It considers an ensemble of $N$ motionless, identical, two-level atoms located in a region of space much smaller than the transition wavelength $\lambda$ and addresses the question of the collective eigenmodes of the system and their radiation rates. In particular, Dicke showed that if all atoms are initially in their excited state, the deexcitation of the ensemble would follow the cascade of all symmetric states, with an acceleration of the emission rate and a coherent emission. All other, nonsymmetric states would never be populated, because they are decoupled from the ladder of the symmetric states. Among these nonsymmetric states, some have an emission rate smaller than the one of a single atom (it can even vanish) and are thus called `subradiant' \citep{Freedhoff:1967}. Note that the symmetry argument relies on hypotheses that were later revealed to be unjustified \citep{Friedberg:1972,Gross:1982}.

A hand-waving argument to describe superradiance could be the following: since all excited atoms are very close to each other, the first, spontaneously emitted photon is immediately `felt' by the other atoms and triggers their stimulated emission, hence a global, collective coherent emission. What is computed in Dicke's paper, among other things, is the dynamics of this emission, with in particular the emission rate at each level of the superradiant cascade. It should be noted that the coherence of the excitation is generated spontaneously, by the first spontaneous emission, which justifies the name of `quantum superradiance' by some authors, e.g. \citet{Berman:2010}, while others have proposed `superfluorescence' \citep{Bonifacio:1975,Gibbs:1977}.

The assumption of a very small size (compared to $\lambda$) may be relevant to microwave spectroscopy, but not so much in the optical domain. To go beyond this `Dicke limit' one has to take into account the distance-dependent dipole-dipole interaction between atoms \citep{Stephen:1964}, as well as propagation effects. With two atoms the whole problem can be solved analytically \citep{Lehmberg:1970b, Stroud:1972, Milonni:1974} but with many atoms in a large sample the situation is much more intricate and was addressed by many theoretical studies in the 1960s-1970s \citep[see, e.g.,][to name a few]{Freedhoff:1967,Ernst:1968, Ernst:1969, Lehmberg:1970a, Arecchi:1970, Friedberg:1971, Rehler:1971, Saunders:1973a, Saunders:1973b, Bonifacio:1975, Ressayre:1976, Ressayre:1977} but experiments were not performed until the 1970s.


\subsection{Experiments in the 1970s-1980s}

Several groups observed and studied superradiance/superfluorescence using atomic vapors or beams, approximately between the mid-1970s and the mid-1980s. References to some of the first observations are \citet{Skribanowitz:1973,Gross:1976,Gibbs:1977} and very interesting reviews are \citet{Feld:1980}, and, mainly for the theoretical aspects, \citet{Gross:1982}.

In almost all the works done in that period, the ideal situation of the Dicke limit is not reached: the samples are large, with often a large aspect ratio (`pencil-shaped'), and propagation/geometrical effects are important \citep[see, e.g.,][but it is only one among many theoretical and experimental studies on those aspects]{MacGillivray:1976}. In that case, the parameter of the sample that governs the superradiant decay rate is related to the optical thickness \citep[see, e.g.,][]{Friedberg:1976}: it is the density of excited atoms times the size along the considered propagation direction.

With this propagation picture in mind it is natural to make a link between superfluorescence and the amplified spontaneous emission (ASE) well known in laser physics \citep{Schuurmans:1979}. In the review by \citet{Feld:1980} it is stated that \textit{``superradiant emission and stimulated emission in a high gain medium are the transient and steady-state forms of phase-coherent amplification, respectively''}, although one can distinguish the limit of a negligible dephasing rate between the dipoles (superfluorescence, with the buildup of a large macroscopic coherent dipole) from the opposite limit of a high dephasing rate, in which ASE and coherent emission can still take place \citep{Schuurmans:1979, Malcuit:1987}.

Finally, the group of Pierre Pillet investigated the possibility of observing subradiance by using a clever choice of three-level atoms \citep{Crubellier:1980} and they succeeded in obtaining some indirect signatures of subradiant states \citep{Pavolini:1985}. This was, to our knowledge, the only observation of subradiance in a many-atom experiment.\footnote{The two-atom case has been studied experimentally with two ions, demonstrating superradiant and subradiant decay rates \citep{DeVoe:1996}. However the change compared to the natural decay rate was only on the order of 1\% because of the relatively large distance between the ions. Other signatures of subradiance in two-particle system have been reported in \citet{Hettich:2002, Filipp:2011, Takasu:2012, McGuyer:2015, Trebbia:2022}.}

\subsection{Matter wave superradiance}

A first revival of the subject took place at the beginning of the ultracold-atom era, with very nice experiments in Ketterle's group using Rayleigh scattering \citep{Inouye:1999,Schneble:2003} and Raman scattering \citep{Inouye:2000,Schneble:2004}. Because of the subrecoil temperature and of the coherence of the atomic sample in these experiments, there is an intricate relationship between the dynamics of the optical field and of the matter wave. In these experiments the atoms are driven by a far detuned field such that there is almost no excited atoms. The similarity with Dicke's superradiance comes from the \emph{momentum} states of the atoms. The atoms in the BEC, which are initially all in the same momentum state, act like a fully inverted systems, and the superradiant dynamics takes place for the transitions between the momentum states driven by the recoil induced by light scattering. From the point of view of the light dynamics, the atomic sample acts as an amplifying media, the gain being provided by recoil-induced resonances.

Several groups performed experiments in this spirit \citep{Yoshikawa:2004, Bar-Gill:2007, Hilliard:2008}, sometimes in combination with a high-finesse optical cavity to enhance the coupling between light and atoms \citep{Slama:2007,Bux:2011,Kessler:2014}. Recently, a form of matterwave subradiance has been observed in the Tübingen group \citep{Wolf:2018}.



\subsection{Single-photon superradiance}

A second revival of the topic was initiated by Marlan Scully and collaborators in 2006 with a paper that addresses what happens when one prepares an ensemble of two-level atoms by absorbing \emph{one} photon of wave vector $k_0$ without any induced optical coherence: there is one excited atom but we don't know which one. In a quantum framework, which is necessary because a single photon has been sent, the atomic sample can be described by the collective entangled state
\begin{equation}\label{eq.TDS}
|\Psi\rangle = \frac{1}{\sqrt{N}} \sum_j e^{i \bm{k_0} \cdot \bm{r}_j} |g_1, g_2,..., e_j,..., g_N\rangle \, .
\end{equation}
The $\bm{r}_j$ are the atoms' random positions and $|g_j\rangle$, $|e_j\rangle$ are the ground and excited states respectively. The phase factor can be interpreted as due to the different possible arrival times of the photon on the atoms, hence the title of the article: \textit{``Directed Spontaneous Emission from an Extended Ensemble of $N$ Atoms: Timing is Everything''} \citep{Scully:2006}. Then one can compute how this collective state decays by spontaneous emission, and one finds that the photon is emitted in the $\bm{k_0}$ direction, which might seem counterintuitive for \emph{spontaneous emission}. Note that there is no condition on the atomic density, the distance between atoms does not have to be small.
However, the state (\ref{eq.TDS}) assumes that the absorption probability was the same for all atoms, which is true only when the optical thickness of the medium is very low (no attenuation of the excitation beam during the crossing of the cloud, and no phase shift either). This paper immediately triggered a lot of discussions \citep{Eberly:2006,Mazets:2007,Scully:2007}. The collective state (\ref{eq.TDS}) has been called the Timed-Dicke (TD) state and the whole subject `single-photon superradiance' \citep{Scully:2009b,Scully:2009}.

The forward emission is however not really surprising, as it is obviously due to the phase-matching condition imposed by the phase factor, which is very similar to the `spin wave' picture commonly used with two-photon Raman transitions in the context of quantum memories \citep{Hammerer:2010,Sangouard:2011}. One can also understand it like an $N$-slit interference effect, where the atoms plays the role of the slits, like in \citet{Grangier:1985}. These analogies are duly mentioned in \citet{Scully:2006}.

Since it is, after all, just an interference effect, it is the same with a continuous driving by a coherent field \citep{Courteille:2010}. Indeed, we know that the outcome of a double-slit experiment is the same with many single photons sent one by one and with a classical field. Although it is less known, it was, actually, already mentioned in Dicke's original paper that the radiation by a coherently-illuminated large sample was in the forward direction \citep[Section \textit{``Radiation from a gas of large extent''} in][]{Dicke:1954}. Of course, with a continuous illumination, the entanglement aspect of the TD state is not relevant any more as it comes from an artificial truncation of the Hilbert space \citep{Eberly:2006,Bienaime:2011}. To our knowledge, only one experiment was realized with a true single-photon source \citep{Frowis:2017}, with the emphasis on the entanglement aspect. For all other experiments, `superradiance in the linear-optics regime' is a better term than `single-photon superradiance'.

A less obvious aspect was predicted slightly later, it is the lifetime (or the rate of collective spontaneous emission $\Gamma_N$) of the TD state, which increases with the number of atoms according to (for $N$ large)
\begin{equation}
\Gamma_N \simeq C \frac{N}{(kR)^2} \Gamma_0 \, ,
\end{equation}
where $R$ is the sample size, $\Gamma_0$ is the spontaneous emission rate for a single atom and $C$ is a numerical factor depending on the geometry. It is therefore indeed a temporal superradiance effect, and not only a directional effect, with an amplification factor which is neither simply the number of atoms (as in the Dicke limit of a small sample), nor the density. In fact, for a spherical sample, $N/(kR)^2$ is proportional to the optical thickness on resonance. This calculation was carried out by several authors following the paper by \citet{Scully:2006} \citep[see][]{Mazets:2007, Svidzinsky:2008, Svidzinsky:2010, Courteille:2010, Friedberg:2010, Prasad:2010}, but we can also find this result in old articles \citep[e.g.][]{Arecchi:1970, Rehler:1971}. Note that the calculation involves taking into account the dipole-dipole interaction between the atoms, so this is really an $N$-body effect and not only an interference effect. It has also been shown that the atoms could be treated as classical dipoles \citep{Svidzinsky:2010}, such that actually the whole problem is purely classical \citep[see also][]{Ruostekoski:1997,Javanainen:1999}. This model has been called the coupled-dipole (CD) model and has been extensively used in the recent years.

Therefore, it appears that the relevant parameter for single-photon superradiance is the on-resonance optical thickness, actually like for superfluorescence experiments and also matterwave superradiance. However, it should be noted that it is much less intuitive, because the picture of a spontaneously emitted photon being amplified by other excited atoms as it propagates through the sample does not work any more. Actually the very existence of cooperative scattering in this regime of linear optics and dilute sample is not intuitive at all, and therefore was not well known.
It is the merit of \citet{Scully:2006} to have brought this subject up to date, at a time where large samples of cold atoms and single-photon sources are available experimentally.
The lack of experiments on superradiance in this regime of linear-optics and large dilute samples was, to our opinion, the reason why it was not well known.
This is why we started an experimental program on this topic, starting with the goal of observing subradiance.

\section{Subradiance in dilute samples in the linear-optics regime}\label{sec.subradiance}

\subsection{Predictions from the coupled-dipole model}

The first predictions of subradiant decay computed from the coupled-dipole model were published in \citet{Bienaime:2012}. A more complete analysis, in particular for the scaling law as a function of the resonant optical thickness $b_0$, was given in the Supplemental Material of \citet{Guerin:2016a}. More details on the numerical simulations were given in \citet{Araujo:2018}. Finally, the comparison between the predictions of the scalar and vectorial CD models was published more recently \citep[see section \ref{sec.vectorial} and ][]{Cipris:2021b}. This last study confirms, \textit{a posteriori}, that the scalar model, which neglects the near-field terms, is an excellent approximation up to densities $\rho_0\lambda^3 \sim 5$, which is well above the typical parameters of our experiment ($\rho_0\lambda^3 \sim 10^{-2})$. Here we consider a Gaussian density distribution and $\rho_0$ is the peak density, $\lambda$ is the wavelength of the transition.

\begin{figure}[ht]
\centering\includegraphics{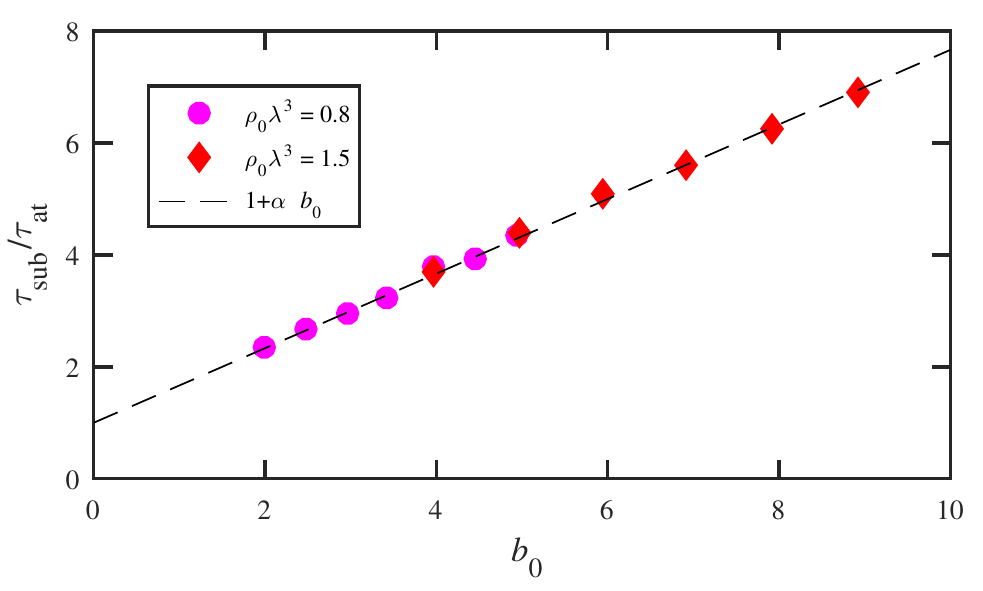}
\caption{Subradiant lifetime as a function of the resonant optical thickness, for two different densities. The lifetime is measured on the decay of the scattered light in a range of intensity (relative to the steady-state level) of $I/I_0 =[10^{-6}, 5\times 10^{-6}]$. Here the scalar model is used with a detuning $\Delta = -15\Gamma_0$. The linear fit gives $\alpha \approx 0.66$. Adapted from \citet{Cipris:2021b}.}
\label{fig.sub_th}
\end{figure}

The main result of these studies is that the late-time decay of the scattered light, after the switch-off of the driving laser, has a characteristic lifetime that scales linearly with the resonant optical thickness $b_0$, independently of the laser detuning and of the density (for $\rho_0\lambda^3 < 5$).  This behavior is illustrated in Fig.\,\ref{fig.sub_th}. Moreover, at large detuning, the whole decay curve becomes independent of the detuning, as discussed in \citet{Guerin:2017a}.

The crucial role played by the resonant optical thickness $b_0 \propto N/(kR)^2$ in super- and subradiance can be understood in different ways. It appears naturally in the optical interpretations discussed in Section \ref{sec.optical}. Another physical explanation is to point out that $N/(kR)^2$ is the ratio between the atom number and the number of distinguishable electromagnetic modes coupled to the sample if $kR\gg 1$ \citep{Guerin:2017a}, and thus acts as the cooperativity parameter. However, the \emph{linear} trend of the subradiant lifetime is only an empirical and approximate numerical result. Due to the numerical limitation of the CD model, it has not been tested at very large $b_0$.

\subsection{Experimental requirements for the observation of subradiance}

The principle of the experiment is quite simple. The atomic sample is made from a standard magneto-optical trap (MOT). We drive it with a probe beam of detuning $\Delta$ for a few microseconds (enough to reach the steady state), switch it off abruptly, and monitor the decay of the scattered light. We use a beam much wider than the cloud such that it can be approximated by a plane wave. 

There are, however, two technical difficulties to do so, a small one and a more challenging one.

The small technical difficulty is to get a fast extinction with a very good extinction ratio for the probe beam. How fast should it be? The timescale of the problem is given by the lifetime of the excited level $\tau_\at = \Gamma_0^{-1} \simeq 26$\,ns, so the switch-off duration should be very short compared to that if one wants to reach the limit of an infinitely fast switch-off. However it is not known \textit{a priori} how critical this criterium is, and for subradiance we are interested in the slow decay visible at late time, so it should not be too critical. We shall come back on this question later (section \ref{sec.sub_multiscat}). For simplicity we only used acousto-optical modulators (AOMs).
Since the switch-off duration in AOMs is given by the sound velocity and the size of the beam, at first sight, one just has to focus stronger to switch off faster. However at some point the Rayleigh length becomes smaller than the crystal length and the size of the beam on the edges of the crystal matters. Moreover it becomes harder to separate the diffraction orders because of the beam divergence, and we want to avoid any spurious light. A strong focusing also spoils the diffraction efficiency because of the associated angular distribution. As a consequence there is a delicate trade-off between the diffraction efficiency, the extinction ratio and the extinction duration. For the extinction ratio, there is always a bit of scattered light on the crystal surface that can enter the following optical path, even through a single-mode fiber, so we used a second AOM in series. After a very careful optimization we got a switch-off duration of $\sim 15$\,ns (defined as the fall time between the 90\% and the 10\% levels) and an extinction ratio better than $10^4$.

The challenging difficulty is to be able to \emph{measure} a fast and good extinction. At first sight, one could think that any fast detector could do the job. Alas, this is not so simple. Even with a high-bandwidth, low-noise detector, the measured switch-off is fast until it reaches the level of typically $\sim 1-2\%$ of the initial steady-state level, and then there are some slow relaxation and/or spurious ripples, which can last for a very long time. Working in the photocounting regime does not, in general, solve the problem because of a phenomena called `afterpulsing': a detected photon has a small probability, on the order on 1\%,  to create a second, fake detection some time later. Although the physical mechanism is different for photomultipliers (PMs) and for avalanche photodiodes (APDs), it exists in both. Then the photons detected just before the switch-off create a fake response after the switch off, at the percent level and for as long as a few microseconds with the detectors we tested.

The solution to this problem was the finding of a new detector, from the company Hamamatsu. Their so-called `hybrid photomultiplier' (HPM\footnote{\url{https://www.hamamatsu.com/eu/en/product/optical-sensors/pmt/hpd.html}}) uses a hybrid technology with a first stage made of a photocathode and a high-voltage electron acceleration, like in a PM, and a second stage made of a semiconductor in which an avalanche process takes place, like in an APD. Somehow, this detector does not have any measurable afterpulsing, and is the only one with this remarkable property, to our knowledge, along with superconducting nanowire single-photon detectors \citep{Zadeh:2021}, which are in a completely different category in term of price, size and complexity. Compared to APDs, it also has the advantage of having a large sensitive area, which is useful for collecting light from a MOT. Thanks to this detector we can measure a clean extinction down to at least $10^{-5}$.


Once those technical problems are solved, the observation of subradiance is straightforward. It only requires some patience because the necessary integration is quite long, on the order of \mbox{500\,000} experimental cycles, corresponding to one full night of integration. Therefore, acquiring a complete set of reliable data taken in the same conditions requires several weeks.

\subsection{Observation of subradiance}

Our first observations of subradiance have been published in \citet{Guerin:2016a} and a follow-up study on the impact of multiple scattering has been reported in \citet{Weiss:2018}. We will come back on the question of multiple scattering in section \ref{sec.sub_multiscat}.

\begin{figure}[ht]
\centering\includegraphics[width=\textwidth]{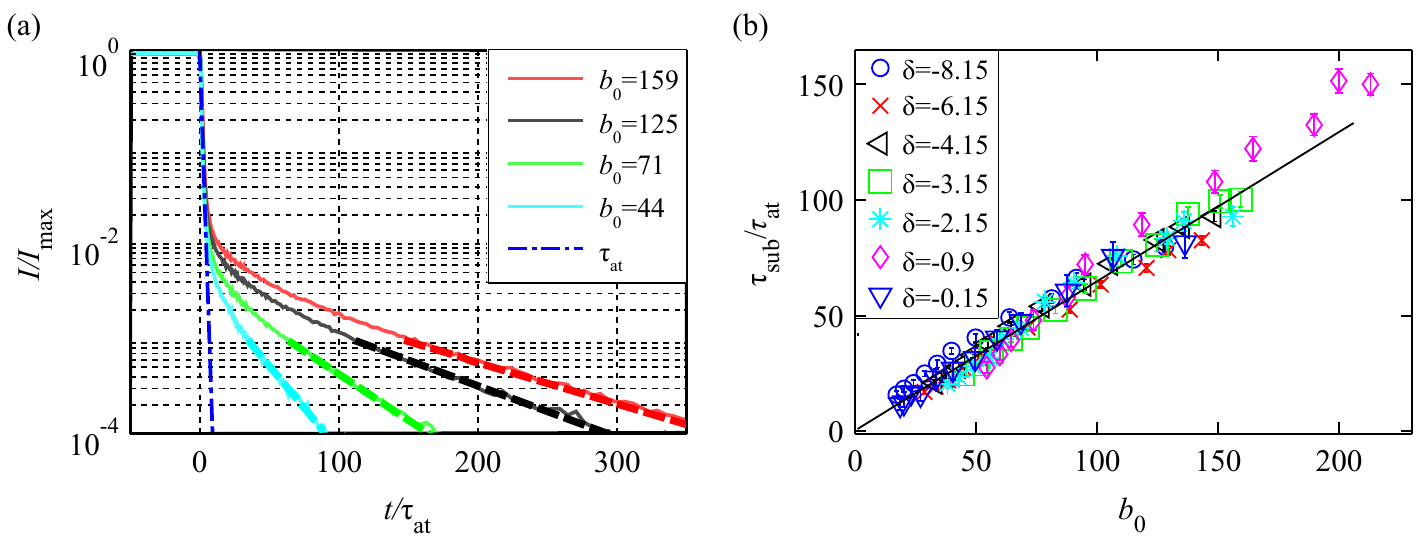}
\caption{(a) Experimental decay curves for different $b_0$, measured with a normalized detuning of $\delta=-3.15$ (in unit of $\Gamma_0$). All curves are normalized to the level right at the switch off of the probe beam. For comparison, the theoretical single atom decay $\tau_\text{at}$ is also shown (dash-dotted line). The slowest decay time $\tau_\text{sub}$ is determined by an exponential fit (dashed lines) at late time. (b) Measured subradiance decay times $\tau_\text{sub}/\tau_\text{at}$ as a function of $b_0$. All data points lie on a single line, independent of the detuning. The linear scaling of $\tau_\text{sub}$ with $b_0$ is stressed by the linear fit (solid line). Reproduced from \citet{Weiss:2018}.}
\label{fig.sub_exp}
\end{figure}

In Fig.\,\ref{fig.sub_exp} we show the main result, which is the subradiant decay and the corresponding fitted lifetime as a function of $b_0$. We observe very clearly what was expected from the CD model, namely a slow decrease of the fluorescence at late time, with a typical lifetime increasing linearly with $b_0$ and independent from the laser detuning.




\subsection{Robustness against the temperature}

One puzzling observation is that, apparently and luckily, the finite temperature of the sample is not a problem for subradiance. This is not intuitive because the at the typical temperature of the experiment, $T \approx \SI{100}{\micro \kelvin}$, the atoms move by more than a wavelength during the subradiant lifetime (equivalently, the Doppler width is larger than the subradiant decay rate). 

\begin{figure}[ht]
\centering\includegraphics[width=\textwidth]{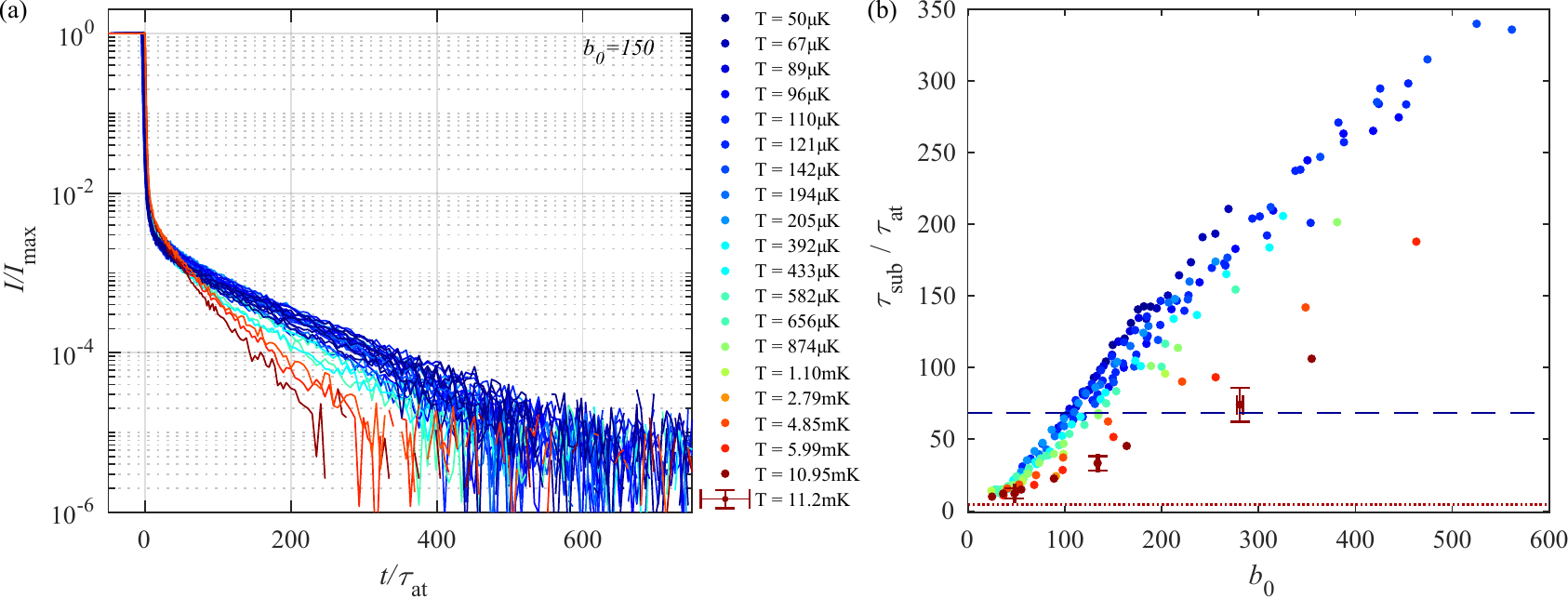} 
\caption{(a) Experimental decay curves for several temperatures, all normalized to the pulse level at the switch-off time $t=0$, with a resonant optical thickness $b_0=150\pm8$. The temperature is encoded in the color scale. A smooth reduction of the subradiant decay with increasing temperature is well visible. (b) Subradiant decay times as a function of $b_0$ for different temperatures (same color code). For clarity only the last data-set is shown with error bars. The horizontal blue dashed line shows the time scale corresponding to the Doppler width for the lowest temperature, \SI{50}{\micro \kelvin}, and the red dotted line the  highest one, \SI{11.2}{\milli \kelvin}. Reproduced from \citet{Weiss:2019}.}
\label{fig.sub_temp}
\end{figure}

This motivated a more systematic study, published in \citet{Weiss:2019}, on the influence of thermal motion on subradiance [Fig.\,\ref{fig.sub_temp}]. As anticipated it shows a nonintuitive robustness, confirmed by CD simulations. The numerics even allow exploring much higher temperature and suggest that subradiance might be visible even at room temperature. One hand-waving explanation of this robustness is that there are so many more subradiant modes than superradiant ones that when a subradiant collective mode is destroyed by atomic motion, there is a high probability that the excitation remains in the subradiant subspace.

Note that another, complementary work has been performed by \citet{Kuraptsev:2020} on the effect of atomic motion on the long-lived state in the dense regime probed near resonance. In that case, corresponding to radiation trapping and Anderson-localized modes, they find that atomic motion can be detrimental.

\section{Superradiance in the linear-optics regime}\label{sec.superradiance}

\subsection{Predictions from the coupled-dipole model}

Whenever there is subradiance, one can expect that superradiance is also there. Although not always true, in the case of a dilute system driven at a large detuning, it is exact. In that case, the system is driven to the Timed-Dicke (TD) state (Eq.\,\ref{eq.TDS}), which is mainly superradiant. 
%
%
%
Indeed, the decay rate of the TD state for a Gaussian cloud is \citep{Courteille:2010}
\begin{equation}\label{eq.Gamma_TD}
\Gamma_\TD = \left(1 + \frac{b_0}{8}\right) \Gamma_0 \,,
\end{equation}
where $b_0 = \sqrt{2\pi}\sigma_0 \rho_0 R$ is the resonant optical thickness at the center of the cloud ($\sigma_0$ is the resonant scattering cross-section\footnote{For a closed transition, $\sigma_0 = 3\lambda^2/(2\pi)$, which is the commonly-used value. However, in the \emph{scalar} CD model, the scattering cross section is actually $\lambda^2/\pi$, which led to some confusion in the first papers of our team, in which the optical thickness was erroneously defined as $3N/(kR)^2$ instead of $2N/(kR)^2$. Correspondingly, the enhancement factor was given as $b_0/12$ instead of $b_0/8$ \citep{Courteille:2010,Bienaime:2010,Bienaime:2011,Bienaime:2012,Bienaime:2013}. This has been corrected in \citet{Bienaime:2014}. Moreover, in the experiment, we don't have a closed transition because of the Zeeman degeneracy. The usual method is to suppose that all Zeeman states are equipopulated, which gives, in the case of the \twothree transition of $^{87}$Rb, an extra factor $g=7/15$. Since the influence of the Zeeman degeneracy on super- and subradiance is not really known, by convention and for consistency with previous papers, we defined $b_0=3N/(kR)^2$ as the `cooperativity parameter', noting that the true optical thickness was $g\, b_0$ in the experiment and $2b_0/3$ in the CD simulations \citep{Guerin:2016a,Araujo:2016,Weiss:2018,Weiss:2019}. More recently, we got convinced that the true optical thickness is really the parameter that governs the physics (see Section \ref{sec.optical}), therefore we should have used that parameter all along.},
$\rho_0$ is the peak density at the center and $R$ is the rms width).

Also, the emission diagram of this collective states is mainly in the forward direction, as discussed in \citet{Scully:2006}. Is superradiance also visible off-axis?
It is tempting to attribute the $b_0$ enhancement term to the forward lobe and the $1$ to the incoherent scattering off axis, which is never completely suppressed because of the finite size of the scattering medium. In that case superradiance would not be visible in the superfluorescence decay off axis.

\begin{figure}[ht]
\centering\includegraphics[width=\textwidth]{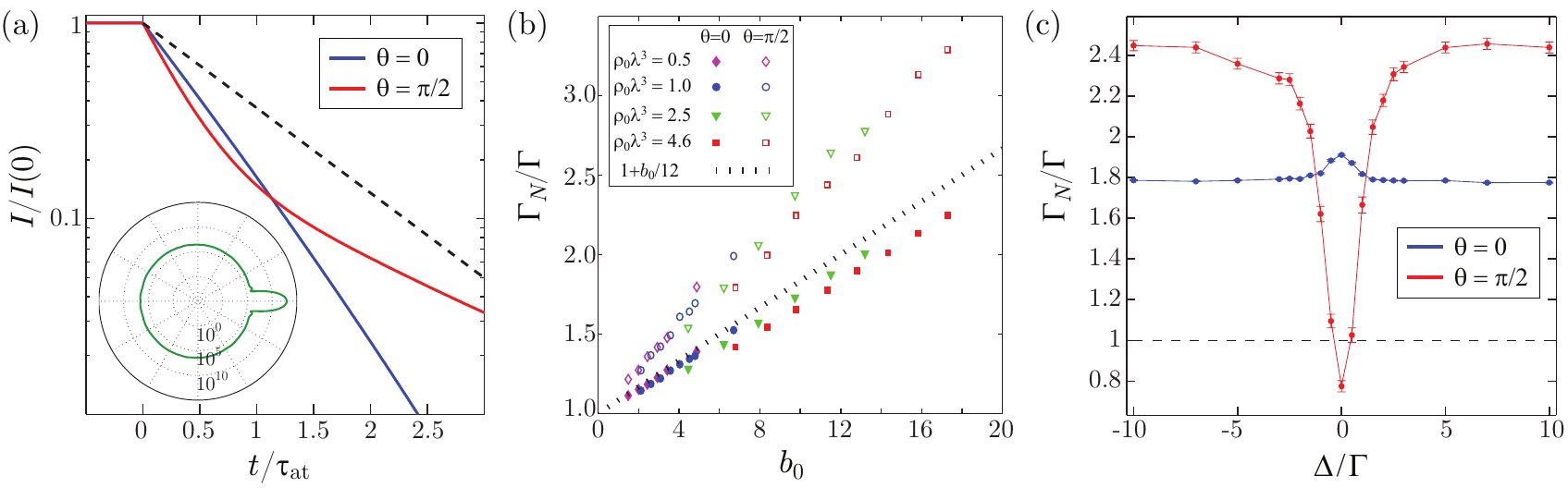}
\caption{Numerical study of the initial collective decay rate $\Gamma_N$. Here we define $b_0 = 3N/(kR)^2$ although we use the scalar model. (a) Temporal evolution of the fluorescence after the switch off of the driving laser at $t=0$, with $b_0=11.3$, $\rho_0 \lambda^3 = 4.6$, $\Delta=10\Gamma$, averaged over 50 configurations, for two different angles, in the forward direction $\theta=0$ and at $90^\circ$ ($\theta = \pi/2$). The amplitude is normalized to the steady state amplitude, which is much larger for $\theta=0$ as shown in the emission diagram (inset, in log scale). An exponential fit in the range $0<t/\tau_\at\leq0.2$ allows us to extract the initial decay rate $\Gamma_N$. The dashed line shows the decay expected for a single atom (rate $\Gamma$). (b) Decay rate as a function of $b_0$ for different densities. Filled symbols are for $\theta=0$ and open symbols for $\theta=\pi/2$. The increase is mainly linear in $b_0$. The slope of the linear increase slightly depends on the angle. The dotted line shows the expectation for the decay of the timed-Dicke state [Eq.~(\ref{eq.Gamma_TD})]. (c) Decay rate as a function of the detuning, for $b_0=17$, $\rho_0 \lambda^3 = 4.6$ and detection angles $\theta = 0, \pi/2$. Off-axis superradiance is suppressed near resonance. Reproduced from \citet{Araujo:2016}.}
\label{fig.sup_th}
\end{figure}

This intuition is however not correct. As shown in Fig.\,\ref{fig.sup_th}, superradiance is also clearly visible off-axis, albeit for a shorter time. The initial decay rate is even larger off-axis.

Another prediction is that off-axis superradiance is suppressed near resonance. One hand-waving argument is that near resonance, the attenuation of the driving field prevents from populating collective modes extended over the whole sample, which is detrimental for superradiance. This argument does not explain why it is different for subradiance. We shall come back on this question later.

\subsection{Experimental observation of superradiance}

In the subradiance data published in \citet{Guerin:2016a}, the superradiant decay was hardly visible because the switch-off was not fast enough. To acquire a new series of data devoted to superradiance, we changed our switch-off setup and replaced the `fast AOM' by an electro-optical modulator (EOM) driven by a pulse generator. The setup is detailed in \citet{Michelle:thesis}. We could reach a falltime of about 3\,ns, and observed superradiant decay rates up to $\sim 6\Gamma_0$.

\begin{figure}[ht]
\centering
\includegraphics[width=\textwidth]{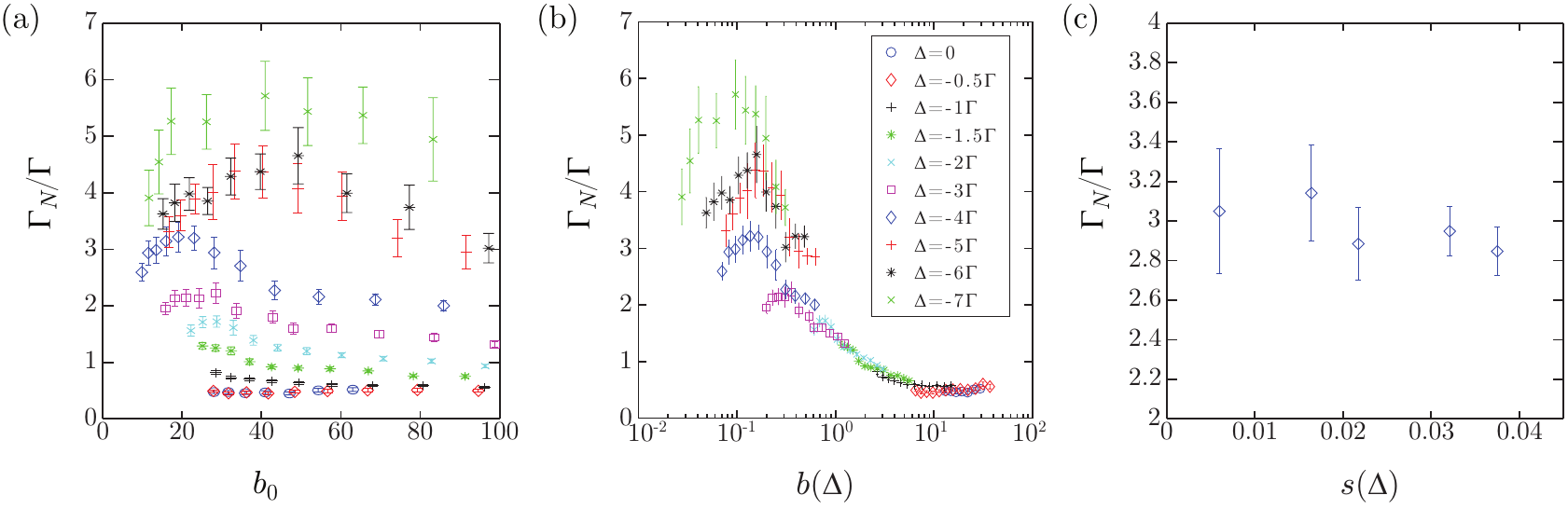}
\caption{Experimental study of the initial collective decay rate $\Gamma_N$. (a) Systematic analysis of $\Gamma_N$ as a function of the resonant optical thickness $b_0$ (still defined as $3N/(kR)^2$) and the detuning $\Delta$. (b) Same data shown as a function of $b(\Delta)$ [Eq.\,(\ref{eq.bDelta})]. When $b(\Delta)\gtrsim 1$, it becomes the scaling parameter. (c) $\Gamma_N$ as a function of the saturation parameter $s(\Delta)$, for $b_0 = 21 \pm 1$ and $\Delta = -4\Gamma$. In all panels, error bars represent the 95\% confidence interval of the fit. Reproduced from \citet{Araujo:2016}.}
\label{fig.sup_exp}
\end{figure}

We report in Fig.\,\ref{fig.sup_exp} the main results of this study. We observe that the collective early decay rate is well superradiant, but not independent of the detuning like for subradiance. Indeed, the decay rates measured for small detunings do not exhibit superradiance, and even at moderate detuning, the decay rate starts to decrease at high $b_0$, when the actual optical thickness
\begin{equation}\label{eq.bDelta}
b(\Delta) = g \frac{b_0}{1+4\Delta^2/\Gamma^2} \,
\end{equation}
is on the order of $1$  or higher. We show indeed in Fig.~\ref{fig.sup_exp}(b) that $b(\Delta)$ becomes the relevant parameter in this regime. These observations are perfectly consistent with the expectation of the coupled-dipole model given at the previous section.

Finally, we checked that the results are independent of the intensity (or the saturation parameter) to confirm that the experiments have been done in the linear-optics regime. For this we varied the intensity $I$ of the probe beam at fixed detuning and $b_0$, and we report in Fig.~\ref{fig.sup_exp}(c) the decay rate as a function of the saturation parameter
\begin{equation}\label{eq.sat}
s(\Delta) = g \frac{I/I_\sat} {1+4\Delta^2/\Gamma^2}\, ,
\end{equation}
with $I_\sat$ the saturation intensity. We observe no significant variation of $\Gamma_N$ with the saturation parameter in the explored range $s<0.04$.

The results were published in \citet{Araujo:2016}, simultaneously to a complementary study by the group of Mark Havey, in which they they investigated the superradiant decay rate in the forward lobe \citep{Roof:2016}. These are, to my knowledge, the first experiments on superradiance in the linear-optics regime. A recent experiment in the case of many excited atoms, and in the dense regime, has been published in \citet{Ferioli:2021b}.


\section{Atomic interpretation of super- and subradiance: collective modes}\label{sec.atomic}

The standard way to describe super- and subradiance, be it in the quantum regime (with many excited atoms) or with classical dipoles, is to consider collective atomic states built from the quantum superposition (or linear combination) of many single atomic states \citep{Dicke:1954}. With classical dipoles it amounts at considering the collective modes of oscillators coupled together via common dissipative channels. The coupling (i.e. the dipole-dipole interaction) lifts the degeneracy of the states such that each collective mode has a slightly different energy (or resonant frequency) and a different lifetime. Then, the modes having a shorter lifetime than a single atom are said to be superradiant, and those with a longer lifetime are called subradiant.


Based on this picture, a widespread method to study collective effects in light-atom interaction is to study the statistical properties of the eigenvalues and eigenvectors of the corresponding `effective Hamiltonian' (which is identical to the matrix of the CD equations). This has been extensively used in particular for the problem of Anderson localization \citep{Rusek:1996, Rusek:2000, Pinheiro:2004, Skipetrov:2014, Bellando:2014, Skipetrov:2015, Maximo:2015, Skipetrov:2016c}.

\begin{figure}[ht]
\centering
\includegraphics[width=0.6\textwidth]{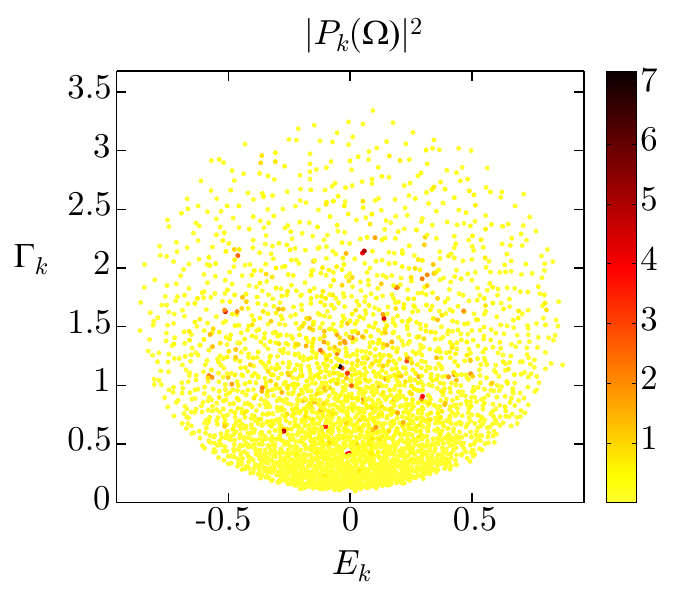}
\caption{Distribution of the eigenvalues $\lambda_k=-\Gamma_k/2+iE_k$ in the complex plane (for $\Delta=0$) with the geometrical factor $|P_k(\boldsymbol{\Omega})|^2$ represented in the color scale. The parameters of the atomic sample are $N=3000$ and $kR\simeq 26.3$, yielding $b_0\simeq 8.7$ and $\rho_0k^{-3}\simeq 10^{-2}$. Reproduced from \citet{Guerin:2017b}.}
\label{fig.population}
\end{figure}

However, in a driven system, the properties of the collective modes are not sufficient to predict or understand the outcome of an experiment. Indeed, the way the system is driven governs which of the collective modes are populated. 
For instance, as discussed in detail in \citet{Guerin:2017b}, it is straightforward to show that the population $p_k$ of the collective mode $k$ is given by
\begin{equation}\label{eq.pop}
p_k = \frac{|P_k(\boldsymbol{\Omega})|^2}{\Gamma_k^2/4 + \left(E_k^0 + \Delta \right)^2} \, ,
\end{equation}
where $E_k^0$ is the eigenfrequency for a detuning $\Delta=0$ of the driving field, $\Gamma_k$ is the linewidth of the mode, and $|P_k(\boldsymbol{\Omega})|^2$ is the projection of the driving field $\boldsymbol{\Omega}$ on the eigenvector $k$ [Fig.\,\ref{fig.population}].

Although this result is intuitive, it explains, at least mathematically if not physically, why superradiance is suppressed near resonance while subradiance is enhanced. Indeed, at large detuning $\Delta$, the denominator is dominated by $\Delta$ and the eigenvalue has little impact on the population, only the geometrical factor on the numerator matters, which favors superradiant states (see Fig.\,\ref{fig.population}). On the contrary, near resonance, the population is enhanced for modes for which $\Gamma_k$ and $E_k$ are both small, i.e. subradiant modes, which is why superradiance is suppressed.

Another limitation of the effective Hamiltonian approach is that we do not measure, in experiments, eigenvalues or eigenmodes. We measure the light emitted from the sample. It can, of course, be computed from the absolute value of the weighted sum of all the decaying eigenmodes, but, because of the nonorthogonality of the eigenvectors, this gives rise to oscillating terms, which may change the dynamics of the decay, even after configuration averaging. It is thus hard to directly relate the statistical properties of the eigenvalues and eigenmodes to experimental observables.

More generally, the coupled-dipole model is very powerful, because it contains all the physics: attenuation, dephasing, dispersion, multiple scattering, diffraction, super- and subradiance, interferences, etc... We argue that this is also a weakness because it does not provide any way to pinpoint the role of the different phenomena, therefore its possibilities for providing useful physical interpretations are limited. That is why it is fruitful, to our opinion, to develop an optical interpretation of super- and subradiance.

\section{Optical interpretation of super- and subradiance}\label{sec.optical}

So far, our modeling and interpretation of super- and subradiance are based on the coupled-dipole model and the associated collective modes. In this picture the light provides an effective interaction between atoms, and we only consider the dynamics of the atoms, supposed to be classical dipoles. Alternatively, it should be possible to understand what happens by considering the light propagating and being scattered inside the disordered atomic medium. The complementarity and equivalence between the two approaches are discussed, for instance, in \citet{Lax:1951,Lagendijk:1996}, although some phenomena may be easier to describe in one framework or the other. It is, for example, easier and more efficient to describe incoherent multiple scattering (radiation trapping) by a radiative transfer approach, such as random walk simulations \citep{Labeyrie:2003,Labeyrie:2005}. Another example is coherent backscattering, an interference effect that can be easily understood by considering the interference between time-reversed multiple-scattering paths \citep{Labeyrie:1999}, whereas it is certainly challenging to understand from the collective modes, although it can be efficiently computed with the CD model \citep{Rouabah:2014}. On the contrary, super- and subradiance are well described by the CD model, and also easy to understand with the simple idea that coupled oscillators build collective modes, as described in the previous section. However, a lot of properties of super- and subradiance are not intuitive in the atomic picture: Why is subradiance enhanced and why does superradiance disappear near resonance? Is the emitted light polarized or depolarized during the superradiant and subradiant decay? What is the frequency of the scattered light during the superradiant and subradiant decay? What happens if the laser switch-off is smoother? What is the difference between the slow decays due to radiation trapping and due to subradiance? Why is subradiance robust against the temperature and against the Zeeman degeneracy? What if the system is driven by a spatially incoherent field? To answer all these questions, an optical description is much more powerful.

The basic ideas behind this optical description are:

\noindent\textit{1-- The problem is not monochromatic.} Indeed, we are studying the temporal dynamics of the scattered light at the switch-off of the driving laser. As a consequence, we cannot consider the external field as monochromatic. On the contrary, we can see the collective decay as a distortion of the input pulse, which enters as a perfect rectangular function of time and exits with a more complex dynamics. This distortion is due to the dispersive nature of the atomic medium.

\noindent\textit{2-- Scattering takes time.} Indeed, there is a time delay associated to a scattering event by a two-level system, called the Wigner time \citep{Wigner:1955}. For two-level atoms it is
\begin{equation}\label{eq.Wigner}
\tau_\mathrm{W} = \frac{2/\Gamma_0}{1+4\Delta^2/\Gamma_0^2}\, ,
\end{equation}
and it has been measured experimentally with cold atoms \citep{Bourgain:2013}. It is the derivative of the scattering phase with respect to the incoming frequency and it corresponds to a group delay.

It is interesting to note that even the case of a single atom is not so trivial. Indeed, if we consider that we drive a two-level atom with a monochromatic field at a detuning $\Delta$, with a weak saturation parameter, until the steady state, and then we abruptly switch-off the field, the decay of the induced dipole will be exponential with a rate $\Gamma_0/2$ (correspondingly, the emitted light will decay at a rate $\Gamma_0$), independently of $\Delta$. This might seem contradictory with the Wigner time delay, which strongly depends on $\Delta$, and in particular vanishes at very large detuning. This is because the Wigner time is only relevant when there is not too much dispersion, e.g. if we send a Gaussian pulse broad enough to correspond to a narrow spectrum, as in the experiment by \citet{Bourgain:2013}. This is not the case with an abrupt switch-off, for which dispersion effects play a major role, as we will see later.

\subsection{Superradiance: single scattering in an effective medium}\label{sec.Sokolov}

Since scattering takes time, the early decay should be dominated by single scattering. We can thus expect to be able to describe superradiance with a process involving only one scattering event. It is then straightforward to propose a model of superradiance based on the dispersion of the medium in which one scattering event takes place.

Similarly to what is done to compute the optical transients at the switch-off or switch-on of a light pulse transmitted through a resonant sample (optical precursors or flashs) \citep{Jeong:2006, Chen:2010, Chalony:2011, Kwong:2014, Kwong:2015}, the first step is to go to Fourier space, apply Beer's law for the light propagation to the scattering center, then apply the atomic polarizability $\alpha(\omega)$, which describes the effect of the scattering event, and then again Beer's law until free space. An inverse Fourier transform allows computing the time dependence of the scattered field, with the effect of the dispersive surrounding medium. Then one can compute the intensity and average over the possible positions for the scattering event.

In the scalar approximation (for simplicity), it gives
\begin{multline}\label{eq.Sokolov}
I_{\bm{k'}}(t) \propto \int d^3\bm{r} \rho(\bm{r}) \, \left| \int_{-\infty}^\infty d\omega E_0(\omega) e^{-i\omega t} \right. \\ \left. \times \exp\left[i\frac{b_0(\bm{r},\bm{k'})}{2}\tilde{\alpha}(\omega)\right] \, \tilde{\alpha}(\omega) \, \exp\left[i\frac{b_0(\bm{r},\bm{k})}{2}\tilde{\alpha}(\omega)\right] \right|^2,
\end{multline}
where $I_{\bm{k'}}(t)$ is the scattered intensity in the direction $\bm{k'}$ as a function of time, $\rho(\bm{r})$ is the density distribution, $E_0(\omega)$ is the Fourier transform of the incident field,
\begin{equation}\label{eq.alphatild}
\tilde{\alpha}(\omega) = \frac{-1}{i + 2(\omega-\omega_\at)/\Gamma} = \frac{i - 2(\omega-\omega_\at)}{1+4(\omega-\omega_\at)^2/\Gamma^2}
\end{equation}
is the dimensionless atomic polarizability and the $b_0$ terms denote the resonant optical thickness through a part of the cloud, from the position $\bm{r}$ into the direction $\bm{k'}$ and from the incident direction $\bm{k}$ to the position $\bm{r}$.

This equation, which we call linear-response (LR) theory\footnote{We called it `linear-dispersion theory' in previous papers \citep{Guerin:2019,Weiss:2021,Asselie:2022} but linear-response theory seems more appropriate.}, is valid for single-scattering only, since there is only one scattering term, and thus it only gives valid results at \emph{early time} after the switch-off, i.e. only for superradiance \citep{Kuraptsev:2017}.

Note also that the average over the scattering positions is done \emph{after} the squared modulus, i.e. on the intensity: the random phase associated to incoherent scattering\footnote{We only consider elastic scattering here, which is well a coherent process. Nevertheless, a random phase appears due to the random position of the scatterer (except in the forward direction), which is why this process is usually called incoherent scattering \citep[see, e.g.,][]{Ishimaru:book}.} and the associated speckle pattern are averaged out. Indeed, what is computed is formally a quantum-mechanical average, i.e., an average over the disorder configurations. Still, it means that superradiance is not related to the interference between light scattered by different atoms. It is only related to the interference between the different Fourier components of the incident field scattered by the atoms and attenuated/dephased by the surrounding effective medium. It is thus mainly a \emph{dispersion} effect. 


In this approach, one can understand the occurrence of a superradiant decay rate by the spectral broadening of the transfer function induced by the larger value of $b_0$: if the transfer function gets broader in Fourier space, the temporal response gets faster. This is also the intuitive explanation given for the flash effect, which can also have a decay rate faster than $\Gamma_0$ \citep{Kwong:2015}.

\begin{figure}[ht]
\includegraphics[width=0.5\textwidth]{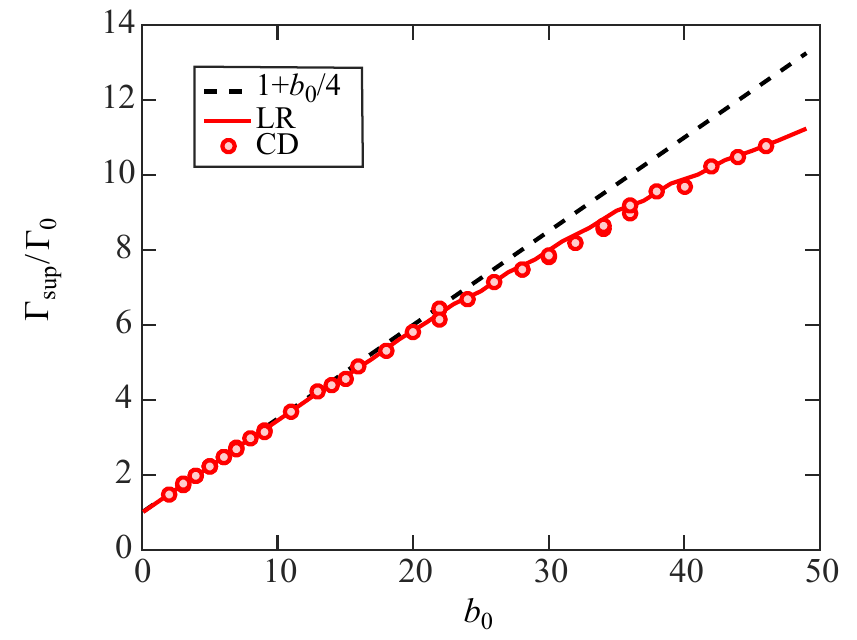}
\includegraphics[width=0.5\textwidth]{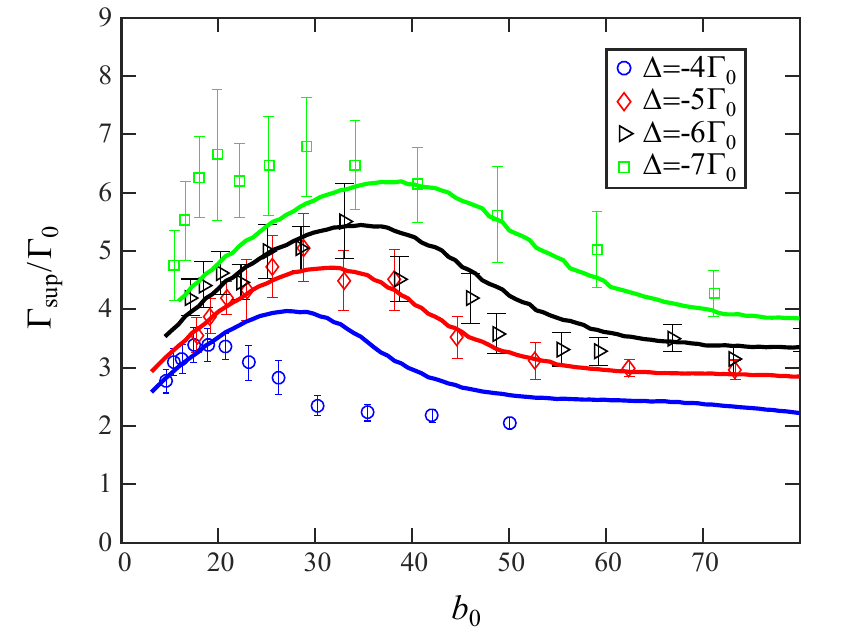}
\caption{Left: Comparison between the coupled-dipole (CD) model and the linear-response (LR) theory for the superradiant decay rate $\Gamma_\supp$. Also shown is the analytical result for the large-detuning limit (Eq. \ref{eq.Gamma_Sokolov}). The detuning is $\Delta=-10\Gamma_0$, the observation direction is $\theta=45^\circ$. The fitting range is $0<t<0.02\tau_\at$. Right: \textit{Ab initio} comparison between the experimental data (symbols) from \citet{Araujo:2016} and the LR theory (solid lines) for the superradiant decay rate $\Gamma_\supp$ as a function of the resonant optical thickness $b_0$ for different detunings. Here the fitting range starts as $t>0.1 \tau_\at$ (to wait for the laser switch-off) and stops when the detected light intensity decreases to $20\%$ from its steady-state value (before the switch-off). Adapted from \citet{Weiss:2021}}
\label{fig.Comparison_LR}
\end{figure}

In Fig.\,\ref{fig.Comparison_LR} we compare the results of the decay rate fitted at very early time on temporal traces computed from the CD model and the LR equation (\ref{eq.Sokolov}), at large detuning ($\Delta=-10\Gamma$). The agreement is excellent. Also shown is the analytical result in the $|\Delta| \rightarrow \infty$ limit, obtained from Eq.\,(\ref{eq.Sokolov}) \citep{Kuraptsev:2017},
\begin{equation}\label{eq.Gamma_Sokolov}
\Gamma_\supp = \left(1 + \frac{b_0}{4} \right) \Gamma_0 \, .
\end{equation}
Note that this does not depend on the observing direction, and that the model only contains superradiance off axis, i.e. with a true scattering event, it does not include the forward lobe of the TD state. As a consequence the decay rate is different, even in the forward direction, than Eq.\,(\ref{eq.Gamma_TD}) for the TD state. The extra $\tilde{\alpha}(\omega)$ term (scattering) is responsible for a factor 2 in the superradiant enhancement factor. It emphasizes the different nature of the forward lobe, which is, in an optical picture, diffracted/refracted light by the effective medium without any true scattering.

A deviation from the analytical result can be seen for the largest $b_0$'s. This is the first sign of the suppression of superradiance as soon as $b(\Delta)$ is not negligible, as observed in \citet{Araujo:2016}. Actually, while doing this comparison, we observed that the good agreement between the two models, as well as the deviation from Eq.\,(\ref{eq.Gamma_Sokolov}), was quite sensitive to the fitting range used for determining the decay rate. The reduction of the superradiant decay rate near resonance is less visible if the decay rate is measured closer to the switch-off time $t=0$. It seems that there is, in fact, always a little bit of superradiance at $t=0^+$, with an unchanged rate, but it is more and more limited in time as the detuning goes closer to resonance, and thus harder to see. This is consistent with a decrease of the superradiant population as discussed in the previous section \citep{Guerin:2017b}.

Of course, this approach is extremely efficient from a computing point of view. Moreover it is not limited in term of atom number or $b_0$, and we can also include the Zeeman structure. Therefore, it is possible to make a direct comparison with experimental data without any free parameters, as shown in Fig. \ref{fig.Comparison_LR}.


Another possible extension is to include the effect of atomic motion, which can be done by a simple Doppler broadening of the atomic polarizability. While the CD simulations with moving atoms are extremely time-demanding \citep{Weiss:2019}, the LR theory allows very fast computing. These results are detailed in \citet{Weiss:2021}.

Finally, note also that the same equation can be used to compute the transient behavior at the switch-on, which exhibits collective Rabi oscillations, which we studied in detail in \citet{Guerin:2019}.

\subsection{Subradiance: multiple scattering of near-resonant light}\label{sec.sub_multiscat}

Although Eq.\,(\ref{eq.Sokolov}) works well for the early decay, it does not capture the full decay dynamics, and in particular the subradiant part. Subradiance is thus due to something that is neglected in the LR theory. One natural idea is multiple scattering, because we know that multiple scattering takes time: the slow diffusion of light in cold atoms, or `radiation trapping' \citep{Holstein:1947, Molisch}, has already been studied in our team \citep{Labeyrie:2003, Labeyrie:2005}. For radiation trapping, the lifetime depends on the actual optical thickness of the sample, i.e. strongly depends on the light detuning. That is why we initially discarded this interpretation of subradiance, whose lifetime is independent of the detuning of the driving field \citep{Bienaime:2012, Guerin:2016a, Weiss:2018}. However, one has to account for the nonmonochromaticity of the problem and consider that there is always a small amount of resonant light, which may dominate the dynamics at the latest times.

\subsubsection{Signatures of multiple scattering in the coupled-dipole simulations}

We show here several signatures of multiple scattering computed from the CD model.
\\

\noindent \textit{Spectrum --}
In the linear-optics regime, there is only elastic scattering, therefore the scattered light has the same optical frequency as the incident light. Nevertheless, once the driving field is off, this frequency is ill-defined. This is because the switch-on/off of the driving field introduces some frequency broadening\footnote{Another way of seeing the same phenomenon is to say that a damped harmonic oscillator will tend to oscillate at its own frequency after the driving force is switched off.}. For in infinitely fast extinction, the corresponding spectrum has broad slow-decaying wings ($\propto 1/(\omega-\omega_\mathrm{L})$ with $\omega_\mathrm{L}$ the laser central frequency).

\begin{figure}[ht]
\includegraphics{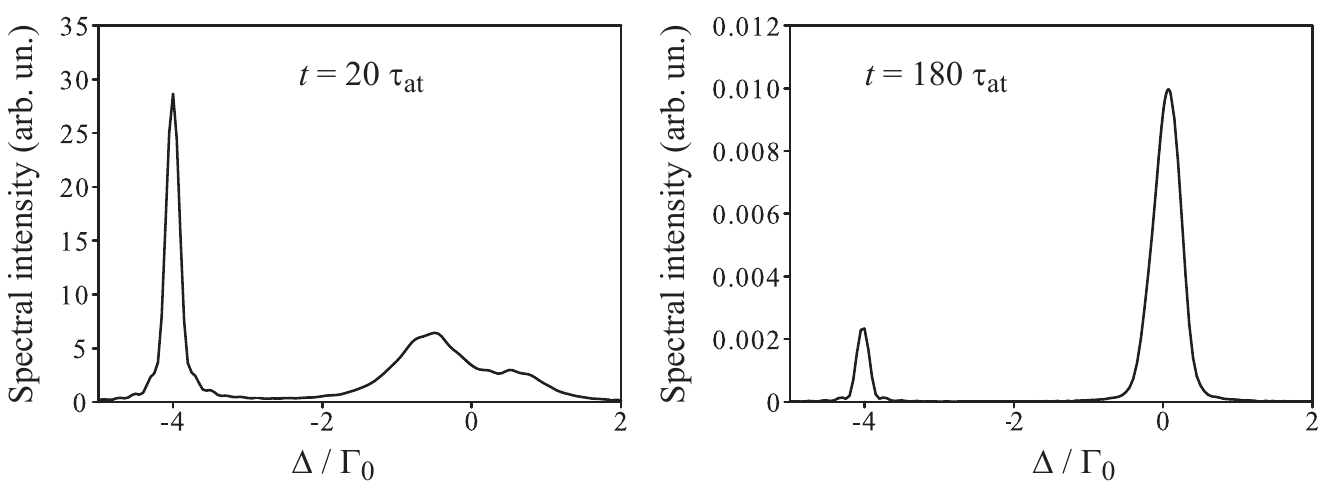}
\caption{Fluorescence spectrum for two time delays after the switch-off of the excitation, $t=20\tau_\at$ (left) and $t=180\tau_\at$ (right). The parameters are $kL=100$  (here, the sample is a cube of size $L$ and homogeneous density $\rho = 0.01 k^3$) and $\Delta=-4\Gamma_0$. Adapted from \citet{Fofanov:2021}.}\label{fig.spectrum}
\end{figure}

From the CD model, one can compute the `instantaneous' frequency of the scattered light by a short-term Fourier transform \citep{Bozhokin:2018}. For Fig. \ref{fig.spectrum} we use a sliding temporal window of width $30 \tau_\at$ and show the spectrum around two delays after the switch-off, $t=20\tau_\at$ and $t=180\tau_\at$. The initial detuning is $\Delta=-4\Gamma_0$.
The figure shows that at later time, the fluorescence is more composed of near-resonant light. This frequency-time correlation is consistent with the idea that long-lived subradiant modes are due to near-resonant light, which is known to undergo multiple scattering.
\\

\noindent \textit{Polarization --}
Another property of the emitted light is its polarization, which can also be computed from the vectorial coupled-dipole model. By driving the system with a circular polarization and computing the light emitted at 90$^\circ$, one can distinguish between single scattering and multiple scattering. Indeed one expects a linear polarization perpendicular to the scattering plane for single scattering, whereas light will be depolarized after many scattering events.

\begin{figure}[ht]
\centering\includegraphics{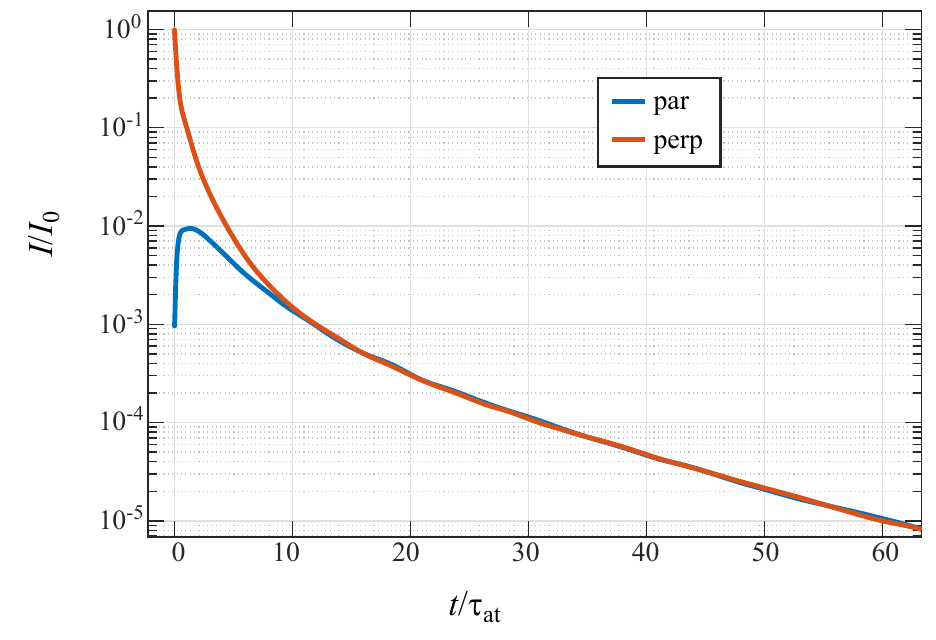}
\caption{Results of CD simulations using the vectorial model. The probe beam has a circular polarization and we compute the light scattered at $90^\circ$, in two orthogonal linear polarization channels: orthogonal (perp) and parallel (par) to the scattering plane. One can see that the superradiant part of the decay is polarized whereas the subradiant part is depolarized. The parameters are $b_0=16$, $\rho_0\lambda^3=5$, $\Delta=-10\Gamma_0$.}
\label{fig.polarization}
\end{figure}

This is what is illustrated in Fig. \ref{fig.polarization}, showing the parallel and the perpendicular (to the scattering plane) polarization channels as a function of time during the decay dynamics. We see that the superradiant part is polarized, whereas the subradiant part is fully depolarized. This is consistent with the interpretation that superradiance is a single-scattering effect and subradiance is due to multiple scattering.
\\

\noindent \textit{Spatial distribution --} Finally, another signature of multiple scattering can be found in the spatial profile of the excitation distribution inside the medium. As shown in \citep{Fofanov:2021}, at late time, this distribution corresponds to the one than can be computed by a diffusion equation, which is a good approximation, at large optical thickness, to describe light transport in the multiple-scattering regime \citep{Labeyrie:2003, Labeyrie:2005}.
\\



\begin{figure}[ht]
\centering\includegraphics[width=\textwidth]{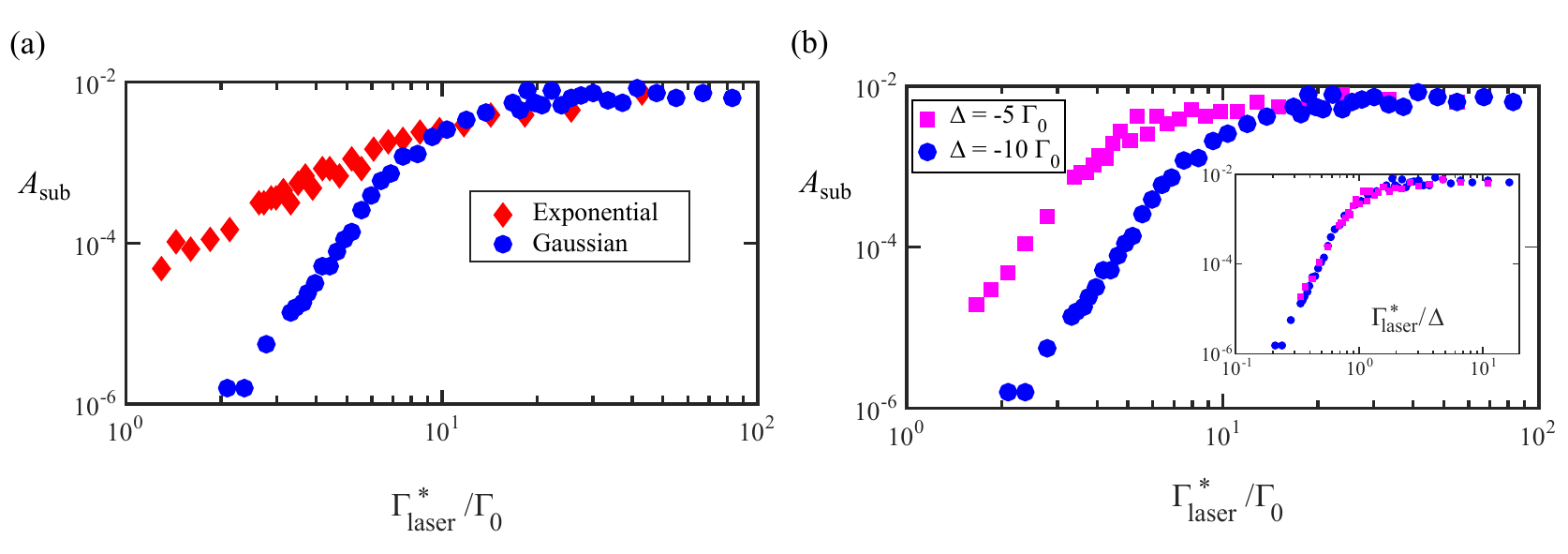}
\caption{Influence of the switch-off profile and rate $\Gamma_\laser^*$ of the driving field on the relative amplitude of subradiance $A_\sub$. (a) Comparison between an exponential and a Gaussian switch-off. The latter case induces less spectral broadening. $\Gamma_\laser^*$ is the width of the spectral broadening corresponding to an exponential or a Gaussian pulse.
The parameters are $b_0=10$ and $\Delta = -10 \Gamma_0$. (b) The switching rate is varied for two different detunings (same $b_0=8.4$), and the inset shows the same data as a function of $\Gamma_\laser^*/\Delta$. From \citet{Asselie:2022}.}
\label{fig.switchoff}
\end{figure}

\noindent \textit{Apodization of the driving pulse --} To test the influence of the switch-off broadening, one can compare the relative amplitude of the slow decay for different switch-off durations and switch-off profiles. This is done in Fig.\,\ref{fig.switchoff}. Here also, the result is unambiguous: the relative amplitude of subradiance seems related to the amount of resonant light. Indeed, at large detuning, an exponential switch-off (inducing a heavy-tailed spectrum) leads to much more subradiance than a Gaussian switch-off (inducing a more compact spectrum), as can be seen in panel (a). In panel (b) we compare the effect of the switching rate for two different large detunings with a Gaussian switch-off: the decrease of the subradiant amplitude for slower switch-off is more pronounced at larger detuning, and in fact, only the ratio between the induced spectral broadening and the detuning matters (inset). This puts in evidence the role of the overlap of the incident spectrum with the atomic resonance. In all cases the time constant is not affected, only the relative amplitude is changed.
\\

All these observations are perfectly consistent with a multiple-scattering interpretation of subradiance.

\subsubsection{Coupled dipoles vs random walk}

If subradiance is due to incoherent multiple scattering, one should be able to simulate it by a radiative-transfer model, such as Monte-Carlo simulations based on a random-walk (RW) process. 

It is straightforward to simulate radiative transport in steady state. The main ingredient is the step-length distribution
\begin{equation}\label{eq.Px}
P(x) = \frac{1}{\ell_\scat} e^{-x/\ell_\scat} \,,
\end{equation}
where $\langle x \rangle = \ell_\scat$ is the mean-free path, computed in the \emph{independent scattering approximation} as $\ell_\scat = 1/(\rho \sigma_\scat)$, with $\rho$ the atomic density and $\sigma_\scat$ the scattering cross-section, which depends on the light detuning $\Delta$ from the atomic resonance:
\begin{equation}\label{eq.sigma_sc}
\sigma_\scat = \frac{\sigma_0}{1+4\Delta^2/\Gamma_0^2}.
\end{equation}
A simple random-walk algorithm allows computing the distribution of the number of scattering event before exiting the medium, and the collective emission diagram of the sample \citep{Chabe:2014, Zhu:2016, Guerin:2017a}. By adding other ingredients, for instance Doppler broadening at each scattering, one can compute other quantities, in that example the emitted spectrum \citep{Eloy:2018}. However there is no intrinsic temporal dynamics in this model.

To recover a temporal dynamics, the standard method, following \citet{Labeyrie:2003,Labeyrie:2005}, is to convert the distribution of the number of scattering events to a temporal trace by multiplying the number of scattering events by the `transport time', which is the sum of two contributions \citep{Lagendijk:1996,Labeyrie:2003}:
\begin{equation}
\tau_\text{tr} = \tau_\text{W} + \frac{\ell_\scat}{v_\text{g}} \, ,
\label{eq:t_tr2}
\end{equation}
with $\tau_\text{W}$ the Wigner time already introduced, and $v_\text{g}$ the group velocity. The three quantities in this expression all depend on the light detuning. However, quite remarkably, the transport time does not \citep[see the Appendix A of][for the detailed computation]{Weiss:2018} and is simply equal to the lifetime of the excited states, $\tau_\at = 1/\Gamma_0$. There is thus no dispersion effect associated to the transport time, only the number of scattering events depends on the detuning. 

In the RW algorithm, all photons enter at $t=0$, the number of scattering events $N_\scat$ is determined from the Monte-Carlo simulation, and then converted to an exit time as indicated: $t_\text{out} = N_\scat \tau_\text{tr} = N_\scat \tau_\at$. Then, to simulate a long square pulse, one has to convolve the result with the pulse itself. That way, it is very hard to take into account the spectral broadening induced by the pulse because it is a nonstationnary process (the spectrum evolves in time). This broadening has been neglected in previous papers \citep{Labeyrie:2003, Labeyrie:2005, Guerin:2016a, Weiss:2018}.

For the sake of comparison with the CD model, one method to overcome this problem is to consider a pulsed excitation with a pulse duration much shorter than $\tau_\at$. Then one can  consider that all photons enter at $t \simeq 0$ with a well-defined  spectrum, which can easily be taken into account in the RW model. This is not rigorous enough to accurately describe the short-time decay (the superradiant part), but it is for the subradiant part, as shown below.

However, using equation (\ref{eq:t_tr2}) is not accurate enough for values of $b_0$ that are not very large. Indeed, the step length distribution (\ref{eq.Px}) is truncated by the finite size of the medium, which gives a shorter effective mean free path, and a \emph{slower} decay at late time. This is somewhat counterintuitive but the explanation is that at late time, the fluorescence is due to near-resonant light, for which the group velocity is negative. Therefore, instead of using $t_\text{out} = N_\scat \tau_\at$, we compute for each photon the escape time
\begin{equation}\label{eq.t_out}
t_\text{out} = N_\scat \tau_\text{W} + \int{\frac{d\ell}{v_\text{g}}} \,,
\end{equation}
where the integral is over the whole random-walk path inside the medium\footnote{Not doing so results in a small but significant difference at late time between the RW and the CD results, as reported in \citet{MonHDR}.}

\begin{figure}[ht]
\centering\includegraphics{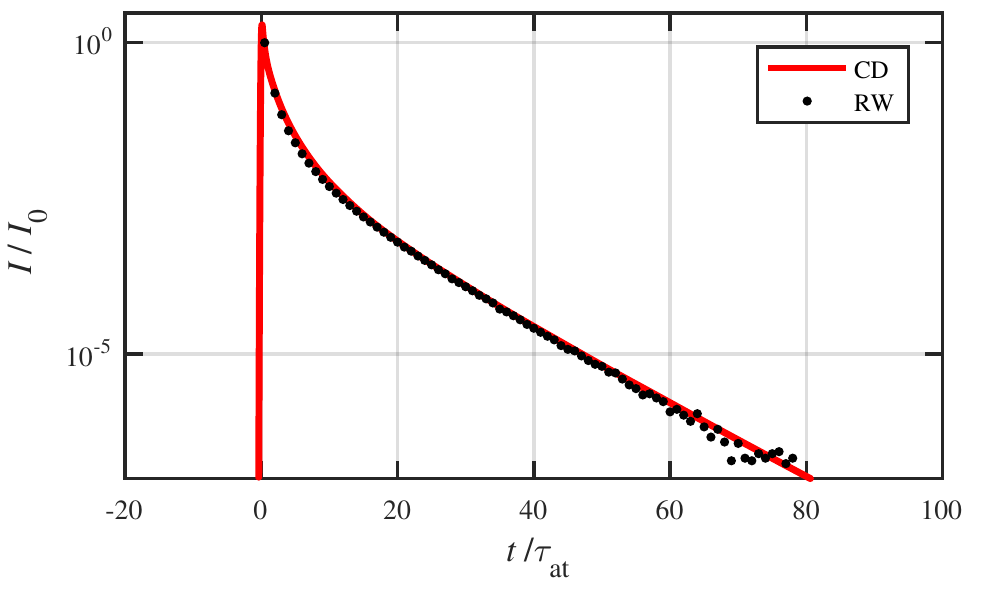}
\caption{Temporal traces of the slow decays following a pulsed excitation computed with the CD and the RW models. The parameters are $b_0=8$ and $\Delta = 10$ (central detuning). The pulse is a Gaussian of rms width $0.1 \tau_\at$ for the field amplitude, inducing a Gaussian spectral broadening of rms width $\sim 7 \Gamma_0$.}
\label{fig.CDvsRW}
\end{figure}

We have performed the comparison between CD and RW simulations in Fig.\,\ref{fig.CDvsRW}, where we show the decay curves computed using the two models. The only adjustable parameter is the vertical scaling (we normalize the integrals). The excitation is a pulse of rms duration $0.1\tau_\at$ at a detuning $10\Gamma_0$ and the resonant optical thickness is $b_0=8$. The agreement between the two models is excellent, which is quite remarkable given their very different nature. This fully validates the multiple-scattering interpretation of subradiance.

\subsection{Influence of the switch-off duration on superradiance}\label{sec.switchoff}

Since the spectral broadening is an essential ingredient of both the LR theory explaining superradiance, and for creating the resonant photons at the origin of multiple scattering/subradiance, it is instructive to study what happens if the switch-off is smoothed. We have already seen that subradiance is suppressed (Fig. \ref{fig.switchoff}), but what happens to superradiance?

\begin{figure}[ht!]
\centering\includegraphics{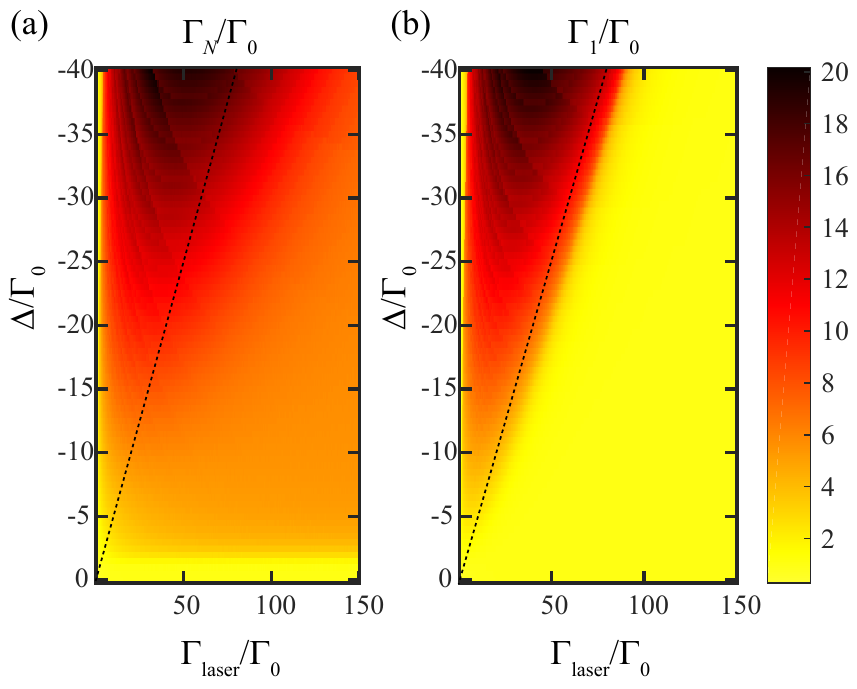}
\caption{Collective decay rate $\Gamma_N$ (a) and single-atom decay rate $\Gamma_1$ (b) as a function of the laser extinction rate $\Gamma_\laser$ and detuning $\Delta$. For panel (a), the resonant optical thickness is $b_0=20$, corresponding to $\Gamma_N^0=6$ for an instantaneous switch-off (Eq. \ref{eq.Gamma_Sokolov}, orange color). The dashed line is $\Delta=\Gamma_\laser/2$, which separates the single-atom physics [above the line, almost identical results in panels (a) and (b)] from the collective physics [below the line in panel (a)]. Reproduced from \citet{Asselie:2022}.}
\label{fig.numerics}
\end{figure}

We use the LR theory in order to investigate the decay dynamics as a function of the switch-off duration of the driving field. We use an exponential extinction profile with a rate $\Gamma_\laser$ and we report in Fig.\,\ref{fig.numerics}(a) the early collective decay rate, $\Gamma_N$, as a function of $\Gamma_\laser$ and the detuning $\Delta$. First, we recover the fact that superradiance is suppressed near resonance (yellow part near the horizontal axis). Second, we obviously observe that with a very slow extinction, the decay simply follows the laser and is slow as well (yellow part near the vertical axis). Third, let us consider a cut at a given, large enough, detuning: the collective decay rate increases as a function of the switching rate of the laser, then reaches a maximum, and then decreases and reaches a plateau. The plateau corresponds to the instantaneous limit, given by Eq.\,(\ref{eq.Gamma_Sokolov}). This nonmonotonous dependence is nonintuitive: it shows that one can achieve a \emph{faster} collective decay rate with a \emph{slower} switch-off compared  with an instantaneous one. This enhancement can be large and is more pronounced for lower $b_0$ and large detuning, as shown in more detail in \citet{Asselie:2022} along with experimental data. The maximum decay rate is obtained near the condition $\Delta \sim \Gamma_\laser$.

To interpret this result correctly, it is instructive to compare with the case of a single atom, as reported in Fig.\,\ref{fig.numerics}(b). It shows that the enhancement of the decay rate with a finite switch-off duration is actually a single-atom effect. The true collective superradiance is only visible for fast switch-off, for $\Gamma_\laser \gtrsim 2\Delta$. The leads us to the following interpretation.

In the limit of a very slow extinction, the decay of the scattered light simply is identical to the decay of the driving field. For larger $\Gamma_\laser$, the system cannot follow the driving pulse any more. One can then see the modified temporal dynamics as a \emph{distortion} of the driving pulse due to its different frequency components. The distortion effect is larger when there are near resonance frequencies because the frequency dependence of the atomic response is much stronger near resonance. 
One can thus minimize the distortion effect, and thus keep a fast fluorescence decay, by minimizing the amount of resonant light with the use of a large detuning and a not-too-fast extinction for the driving pulse. In other words, when $\Delta > \Gamma_\laser > \Gamma_0$, nothing prevents the decay dynamics of the scattered light to be faster than $\Gamma_0$, since there is little distortion.
This is reminiscent of the fact that, at large detuning, the Wigner time delay is very small (see the remark after Eq. \ref{eq.Wigner}).

This description works equally well with a single atom and a large sample. In the latter case, however, a fast decay is recovered even with a larger spectral broadening ($\Gamma_\laser > \Delta$) because the effective medium around the scatterer \emph{filters out} the resonant light by the attenuation in the medium (the exponential terms in Eq. \ref{eq.Sokolov}). This attenuation is not absorption but is due to a second scattering event. The corresponding light is thus delayed in time (by $\tau_\at$ by scattering event in average) and feeds the multiple-scattering orders. As a consequence, the early decay only contains single scattering from off-resonant light and thus appears faster, i.e. superradiant. The larger $b_0$, the wider the frequency filtering, and the faster the decay. 

In light of this new interpretation, it is interesting to reconsider the experimental data. Do we really see the collective superradiance or are we dominated by single-atom physics? A first and easy answer is that the measured decay rates clearly depend on $b_0$ [Figs. \ref{fig.sup_exp}, \ref{fig.Comparison_LR}]. Still, it is interesting to compute what would the decay rates be for the single-atom case with the experimental switch-off profile. We report the computed values of $\Gamma_1$ for the same detunings as in Fig. \ref{fig.Comparison_LR} in Table \ref{tab.Gamma1}: the measured collective decay rates are significantly higher than the single-atom one. We can thus conclude that we do observe a collective superradiance effect, although it is interesting to note that $\Gamma_1>\Gamma_0$ because the switch-off is not infinitely fast and because of the large detuning.

\begin{table}[h]
    \centering
    \caption{Single atom decay rate with the experimental switch-off profile of the superradiant experiment \citep{Araujo:2016, Weiss:2021}.\\}
    \label{tab.Gamma1}
    \begin{tabular}{l|cccc}
        $\Delta/\Gamma_0$ & -4 & -5 & -6 & -7 \\
        \hline
				$\Gamma_1/\Gamma_0$ & 1.5 & 1.9 & 2.3 & 2.7 \\
    \end{tabular}
\end{table}

\subsection{Conclusion on the optical picture}

We have thus arrived to a complete and self-consistent picture of superradiance and subradiance that only relies on classical optical mechanisms such as dispersion and multiple scattering \citep{Weiss:2021, Fofanov:2021, Asselie:2022}.

In summary, it consists in considering the different frequency components of the light that are necessarily present in a time-dynamical experiment. The most off-resonant light undergoes at most only one scattering event and, thanks to a short Wigner time, can escape the medium very fast, after a delay mush shorter than the natural lifetime: it appears superradiant. When the optical thickness increases, this early decay is dominated by light that is farther from resonance, therefore it appears more superradiant. Correspondingly, the subradiant lifetime is increased as it corresponds to the multiple scattering of the near-resonant light.

This alternative description, complementary to the equally-valid collective atomic mode picture, sheds a new light on cooperative scattering in dilute disordered samples and, to our opinion, provides a more intuitive understanding of most properties of superradiance and subradiance that have been observed experimentally or numerically: the increase of subradiance and the disappearance of superradiance when the system is driven near resonance, the fact that the superradiant emission is polarized and the subradiant one depolarized, that fact that the superradiant emission can appear faster while subradiance is suppressed if we apodize the switch-off profile, the spectral properties of the superradiant and subradiant emission, etc. 


A remaining open question is the scaling of the subradiant lifetime with $b_0$. We have numerically observed that it is linear at moderate $b_0$ [Fig. \ref{fig.sub_th}] but we have not find an analytical proof of that, neither within the CD model, nor with the RW model. Moreover, the multiple scattering interpretation suggests that, at very large $b_0$ (and for motionless atoms), the scaling should change as $b_0^2$ (see, e.g., \citet{Labeyrie:2003, Labeyrie:2005, Weiss:2018}). We have not observed this behavior, nor in the CD simulations, probably because of the computing limitations, which prevented us from investigating large enough $b_0$ without density effects, neither in the experiment, probably because of the frequency redistribution due to the Doppler broadening, which breaks this $b_0^2$ scaling \citep{Pierrat:2009}, which has also not be seen in previous radiation-trapping experiments \citep{Labeyrie:2003, Labeyrie:2005} for the same reason.


\section{Beyond the dilute and linear-optics limits}\label{sec.beyond}

So far we only addressed the limit of a vanishing density for the atomic sample and a vanishing intensity for the driving field. In this last section, we investigate the first corrections to these limiting cases. 


\subsection{Influence of pairs}

\begin{figure}[ht!]
\centering\includegraphics[width=\textwidth]{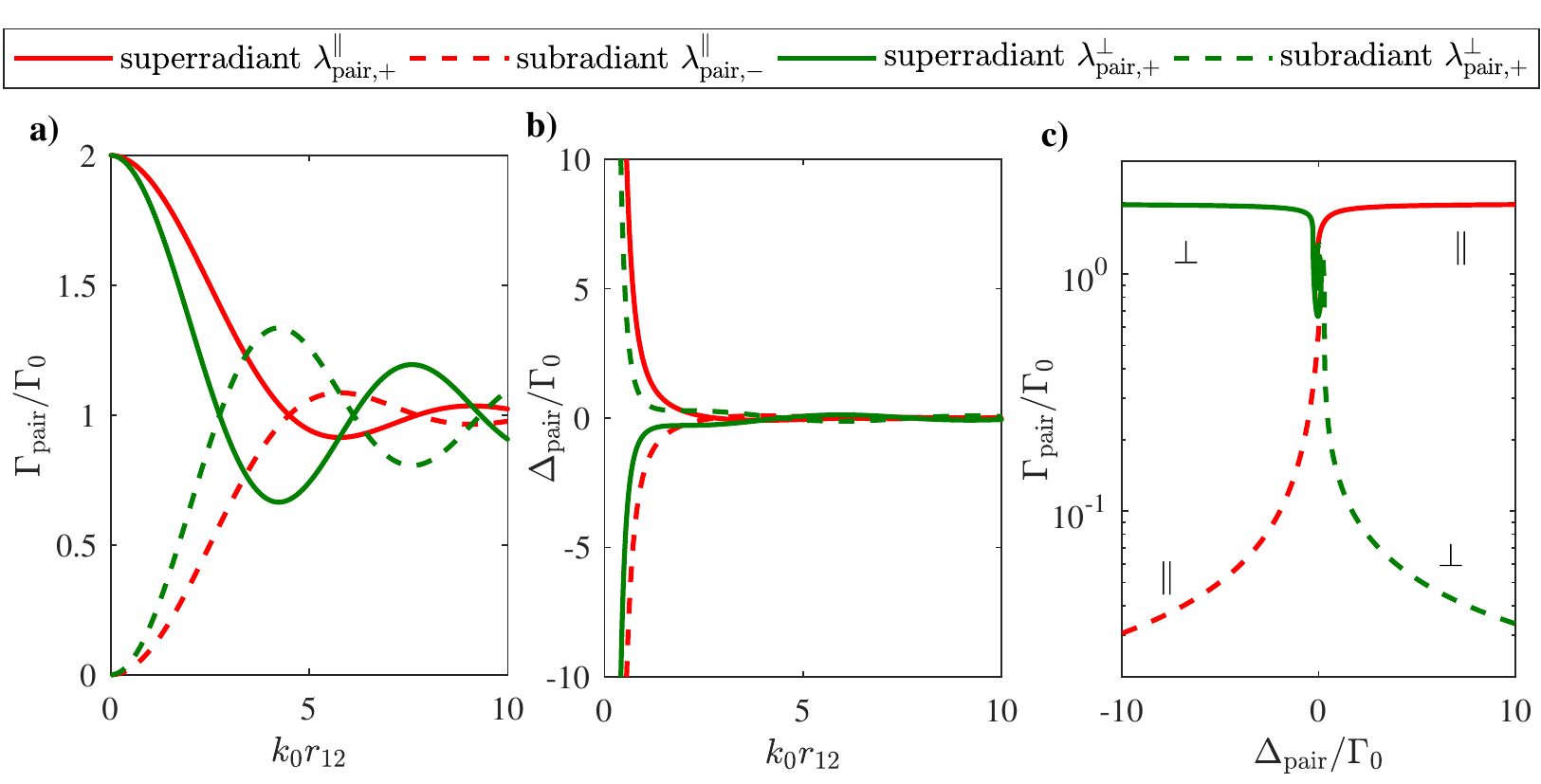}
\caption{Complex eigenvalues $\lambda_\pair = i\Delta_\pair - \Gamma_\pair/2$ corresponding to the modes of atom pair in the vectorial coupled-dipole model with $N=2$ (Eqs.\,\ref{eq.pairs1}-\ref{eq.pairs2}), as a function of the interatomic distance $r_{12}$ (a,b) and in the complex plane (c). Reproduced from \citet{Cipris:thesis}.}
\label{fig.pairs}
\end{figure}

Even at low density, there is still a finite probability that two atoms, which we suppose here to be motionless, are very close together. Due to their strong dipole-dipole interaction, they build a superradiant/subradiant pairs, with strongly shifted resonances $\Delta_\pair$ and modified decay rates $\Gamma_\pair$, as given by the eigenvalues $\lambda_\pair = i\Delta_\pair - \Gamma_\pair/2$ (see Fig. \ref{fig.pairs}):
\begin{eqnarray}\label{eq.pairs}
\lambda_{\pair, \pm}^{\parallel} &=& -\frac{\Gamma_0}{2} \left[ 1 \pm \frac{3}{2i} e^{i k_0 r_{12}} \left( -\frac{2i}{(k_0 r_{12})^2} + \frac{2}{(k_0 r_{12})^3} \right)\right] \, ,\label{eq.pairs1}\\
\lambda_{\pair, \pm}^{\bot} &=& -\frac{\Gamma_0}{2} \left[ 1 \pm \frac{3}{2i} e^{i k_0 r_{12}} \left( \frac{1}{k_0 r_{12}}+\frac{i}{(k_0 r_{12})^2} - \frac{1}{(k_0 r_{12})^3} \right)\right] \label{eq.pairs2}
\end{eqnarray}
The $+$ and $-$ symbols denote the symetric (superradiant) and antisymetric (subradiant) states respectively, while the $\parallel$ and $\bot$ symbol denote the orientation of the dipoles from the interatomic axis. The $\lambda_\pair^\bot$ are doubly degenerate.

Although the relative number of atoms involved in such pairs is proportional to the density, and is thus very small at low density, the absolute number of pairs \emph{increases} when we simultaneously decrease the density and increase the atom number such that the resonant optical thickness is constant. Indeed, for a homogeneous sample of size $R$ containing $N$ atoms at density $\rho$, and if we define a pair as two atoms within a distance $r<1/k_0$, the average number of pairs is $N_\mathrm{pair} \sim N \times \rho/k_0^3$, which gives, as a function of $b_0$ and $\rho$, $N_\mathrm{pair} \sim b_0^3/\rho$.

Because of the large energy shift of these pairs, they can have a relative contribution in the experimental signal much larger than their relative number, if the system is driven with a large detuning close to some pairs' resonance. This effect is discussed in detail by \citet{Fofanov:2021}: pair subradiance can dominate the dynamics at long time in some regime of parameters, even in the dilute limit. A good indicator of the influence of pair is a red-blue asymmetry (see Fig. \ref{fig.pairs}c).

In experiments with moving atoms, it is plausible that the effect of the pairs is strongly reduced, because the atomic velocity prevents the atomic positions to remain fixed, and because the light-assisted inelastic collisions should destroy them \citep{Caires:2004, Fuhrmanek:2012}. 

That is why, in all previous computations with the CD model (except Fig. \ref{fig.spectrum}), an exclusion volume was used in order to only discuss the collective long-lived modes involving many atoms. Nevertheless, the possible influence of pairs in experimental conditions remains to be studied.


\subsection{Density effect}\label{sec.vectorial}

We now address the modifications of subradiance at large density. We do not consider the Dicke limit or the very large density regime $\rho\lambda^3\gg1$, but merely the first correction to the dilute regime, when $\rho\lambda^3\gtrapprox 1$.

\begin{figure}[ht!]
\centering\includegraphics[width=\textwidth]{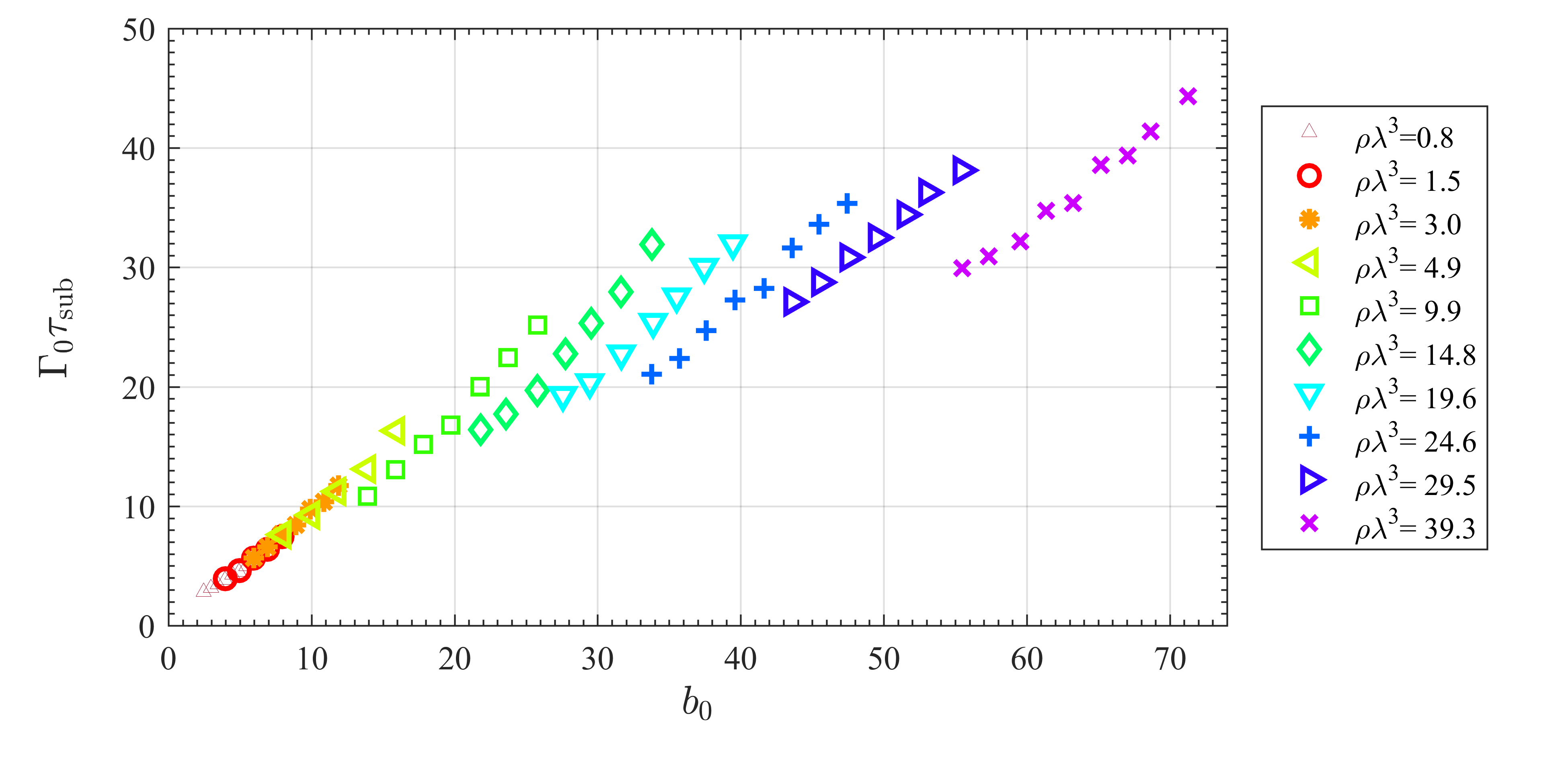}
\caption{Subradiant lifetime $\tau_\sub$ as a function of the on-resonance optical thickness $b_0$ for several densities of the sample. The detuning is $\Delta = -15 \Gamma_0$ and the incident polarization is circular. The total intensity is measured at $\theta = 45^\circ$. Reproduced from \citet{Cipris:2021b}.}
\label{fig.vectorial}
\end{figure}

We have numerically shown in \citet{Cipris:2021b} that the near-field dipole-dipole interaction, which only becomes significant at large density, induces a decrease of the subradiant lifetimes [Fig. \ref{fig.vectorial}], but this decrease remains modest in the range of explored densities. At much higher density, near the Dicke limit, it has been shown that the cooperativity parameter becomes the atom number \citep{Ferioli:2021a}.

By looking at the eigenvalue distribution, one can associate this decrease of the subradiant lifetime with a spreading of the eigenenergies of the collective modes \citep{Cipris:2021b}. One can thus interpret this effect as a decrease of the effective optical thickness due to the inhomogeneous broadening induced by the near-field dipole-dipole interaction.

Note that this is quite similar to what has been observed in the context of Anderson localization of light in resonant two-level systems. There too, the near-field interaction kills localized modes \citep{Skipetrov:2014, Bellando:2014}. This is consistent with the fact that the two phenomena are based on multiple scattering of light.

Finally, note also that a similar near-field--induced broadening has been mentioned as being responsible for the saturation of the refractive index of dense gases \citep{Andreoli:2021}.

\subsection{Quantum regime}

Finally, we have also explored how subradiance is modified beyond the linear-optics regime, i.e. by slightly increasing the saturation parameter $s$ (Eq.\,\ref{eq.sat}) \citep{Cipris:2021a}. Like for the density, we only investigated the first deviation from the regime $s\ll 1$, only using $s \gtrapprox 1$ and not $s \gg 1$.
In this regime, the classical coupled-dipole model is not enough and one has to consider the excited-state populations. 

\begin{figure}[ht]
\centering\includegraphics[width=0.6\textwidth]{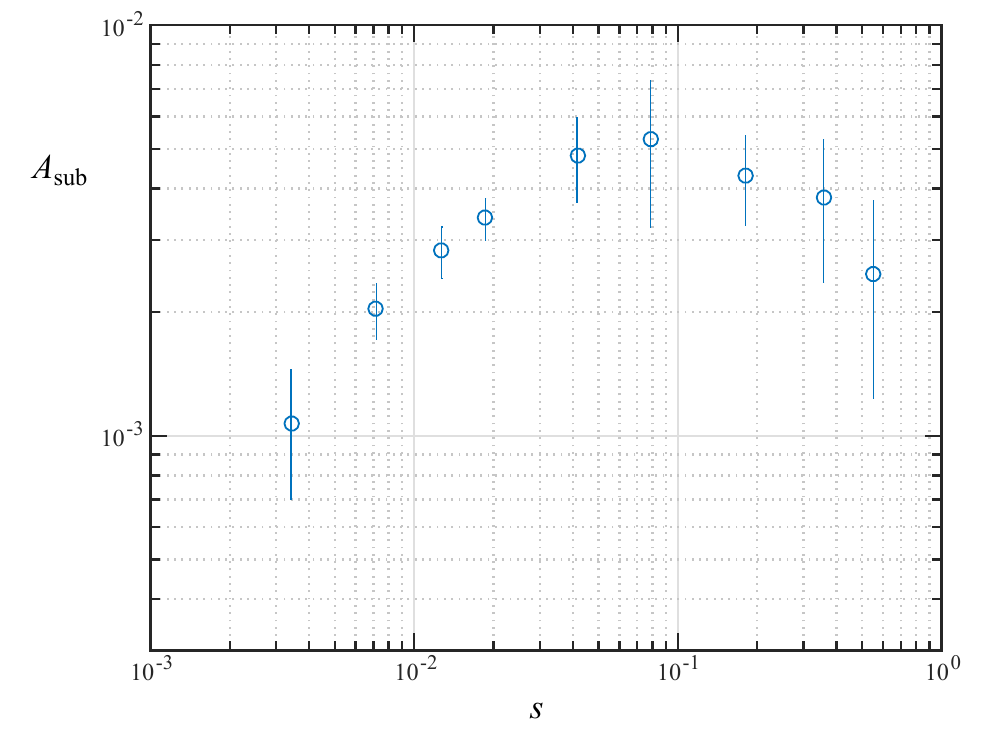}
\caption{Experimental measurement of the relative subradiant amplitude as a function of the saturation parameter of the probe beam. The parameters are $b_0=54$ and $\Delta = -4\Gamma$. Adapted from \citet{Cipris:2021a}.}
\label{fig.Asub_vs_s}
\end{figure}

Our main experimental observation is that the late-time decay rate is not modified, but the relative amplitude of the subradiant part is.
We measure this relative amplitude on decay curves similar to the ones of Fig. \ref{fig.sub_exp}(a), in which we fit the late time decay by an exponential, yielding a lifetime $\tau_\sub$ and an amplitude $A_\sub$. The decay curves are normalized and so $A_\sub$ is the relative amplitude of subradiance. We show in Fig.\,\ref{fig.Asub_vs_s} $A_\sub$ as a function of the saturation parameter $s$.

One puzzling observation is that we do not observe any plateau at low $s$, even going down to $s \sim 3\times10^{-3}$: we do not reach the linear-optics regime! This seems to be in contradiction with Fig.\,3(b) of the Supplemental Material of \citet{Guerin:2016a}, but the range of explored value of $s$ was more limited, and we might have by chance only explored the top of the curve, where it is locally flat. However, this is in line with the prediction of \citet{Williamson:2020}, which suggests that subradiant modes saturate at lower intensity than independent atoms, with an effective saturation parameter scaling as $\Gamma_\sub^{2.5}$. It is thus extremely hard to reach the linear-optics regime with long-lived modes.

The second observation is that the subradiant amplitude increases with the saturation parameter. It can be interpreted in the framework of collective Dicke states by an optical-pumping effect via the multiply-excited superradiant states: as the saturation parameter increases, some population is created beyond the single-photon Dicke states, and those states can partly decay towards the subradiant states. This population adds up to the one created by direct driving from the ground state, which is the linear-optics mechanism. They ultimately populate the single-photon subradiant states, and so the late-time decay rate is similar. This interpretation has also been used for data taken at larger saturation parameters by \citet{Glicenstein:2022}.

These observations have been compared to several models: exact solution with only a few atoms, BBGKY approach truncated to second-order, i.e. including quantum pair correlations, and mean-field approximation, i.e. without any quantum correlations. The observations are qualitatively in disagreement with the mean-field model but in agreement with the exact model and the second-order BBGKY approach, showing that quantum pair correlation plays a role \citep{Cipris:2021a}.

Finally, it is interesting to note that the optical picture of subradiance, namely multiple scattering of resonant light, also provides a nice and simple interpretation of this experiment. Indeed, at any finite saturation parameter, there is a proportion of inelastic scattering. At low $s$, this proportion increases linearly with $s$, and at large $s$, elastic scattering vanishes. Inelastic scattering gives rise to the famous Mollow spectrum \citep{Mollow:1969}, a sideband of which is on resonance with the dressed atom. As a consequence, the proportion of resonant light after one scattering event is proportional to $s$. At very large $s$, the light shift of the dressed atom is significant and stops when the field is switched off, such that the inelastically scattering light is not resonant any more. This can explain the drop of $A_\sub$ observed at large $s$.

The two pictures are consistent with each other. Indeed, inelastic scattering comes from the quantum fluctuations of the dipole, so that quantum pair correlations correspond, in this setting, to the scattering by one dipole of the field created by the quantum fluctuations of another dipole.

A remaining open question is the physical explanation of the observed power-law behavior, $A_\sub \propto s^{1/2}$ at low $s$. The two-atom toy model discussed in \citep{Cipris:2021a} as well as the optical picture would rather predict $A_\sub \propto s$. The second-order BBGKY approximation also obtains $A_\sub \propto s^{1/2}$, but this is only an empirical numerical result and not an analytical one.

\section{Conclusion}\label{sec.conclusion}



We have presented in this Chapter a summary of the work done in our group over several years on superradiance and subradiance measured as a collective temporal dynamics at the switch-off of the driving field \citep{Bienaime:2012, Guerin:2016a, Araujo:2016, Guerin:2017a, Guerin:2017b, Araujo:2018, Michelle:thesis, Weiss:2018, Weiss:2019, Weiss:2021, Cipris:2021a, Cipris:2021b, Cipris:thesis, Fofanov:2021, Asselie:2022}. More details can be found in these references. Since our understanding evolved over time, the presentation is more consistent here (hopefully). Although not discussed here, note that there is also a nontrivial collective dynamics visible at the switch-on \citep{Guerin:2019, EspiritoSanto:2020}.

The traditional way of describing and understanding super- and subradiance is based on Dicke collective atomic states. This is very powerful and versatile, as it can address the single-excitation case as well as many excitations or even a completely inverted system, superradiant and subradiant states, low and large density, ordered or disordered samples, etc. In particular, a large body of theoretical work \citep[and one experimental realization, ][]{Rui:2020,Srakaew:2023} has been published  over the last years on ordered atomic arrays with subwavelength spacing \citep[see, e.g., ][to name a few]{Facchinetti:2016, Bettles:2016b, Bettles:2017, Shahmoon:2017, Asenjo:2017, Perczel:2017, Manzoni:2018, Needham:2019, MorenoCardoner:2019, Bekenstein:2020, MorenoCardoner:2021, Holzinger:2021, Rubies-Bigorda:2023}. On the contrary, most of our own work is focused on large and dilute disordered samples.

Within the linear-optics limit, the coupled-dipole model is also very powerful and can be used to address other problems as well, such as Anderson localization, the properties of ordered mesoscopic ensembles, etc. Nevertheless, even if the properties of superradiance and subradiance can be computed from the CD model, it often does not provide a physical intuition. 

Experimentally, we do not measure eigenvalues or eigenstates, but we shine light onto the atomic sample, and we detect light emitted by, or scattered off, the sample. It is thus natural to ask what the light is doing in between. To answer this question, we have developed an optical description of superradiance and subradiance in the linear-optics regime. They appear as a dispersion effect, the off-resonant component of the light undergoing an accelerated decay due to the short Wigner time, while the near-resonant light undergoes radiation trapping due to multiple scattering. From this simple description, most properties become physically intuitive. In some situations, it is also computationally very efficient.

This optical picture is only valid for 3D disordered samples and for off-axis measurements averaged over the disorder configurations, such that interference effects are averaged out. It is also mainly useful at low density, when one can use the refractive index and the scattering cross-section of a single-atom as building blocks of the model. It is also relatively simple in the linear-optics regime. Beyond the linear-optics regime, one would need to add other ingredients beside dispersion and multiple scattering, such as inelastic scattering and gain. Then an optical model would be quickly intractable, although it should in principle be feasible. 


Some might think that if everything can be explained by classical optics, then it's not so interesting. Well, it is our duty as physicists and researchers to understand things as deeply as possible. Once the concepts are demystified, they might seem obvious: it means that we have done a good job.

\section*{Acknowledgements}

I thank all my students and collaborators for their contribution to this work throughout the years: Michelle Ara\'ujo, Ana Cipris, Stephan Asselie, Patrizia Weiss, Romain Bachelard, Johannes Schachenmayer, Igor Sokolov, and Robin Kaiser. I would like to particularly emphasize the important contributions made by Patrizia Weiss and Ana Cipris, and the very precious insight of Igor Sokolov for the understanding of subradiance. I also thank Ana Cipris for letting me use some of her unpublished figures. In addition, I thank Mathilde Hugbart and Guillaume Labeyrie for the friendly atmosphere in the group and the useful scientific discussions. Last but not least, I warmly thank Robin Kaiser for the numerous, always fruitful discussions, and his constant support and help over the last 15 years.

Part of this work was performed in the framework of the European Training Network ColOpt (EU H2020 program), Grant Agreement No.\,721465, and the project ANDLICA, ERC Advanced Grant No.\,832219. We also acknowledge funding from the French National Research Agency within the projects LOVE (No.\,ANR14-CE26-0032), PACE-IN (No.\,ANR19-QUAN-003) and QuaCor (No.\,ANR19-CE47-0014).

\section*{References}


\end{document}